\documentclass[aip, jcp, reprint, nolinenumbers, twocolumn, nobalancelastpage]{revtex4-1}

\usepackage{todonotes}
\usepackage{soul}
\usepackage{enumitem}
\usepackage{graphicx}          
\usepackage{dcolumn}           
\usepackage{bm}                
\usepackage[mathlines]{lineno} 
\usepackage[utf8]{inputenc}
\usepackage{mathrsfs}
\usepackage[version=4]{mhchem}
\usepackage[]{algorithm}
\usepackage{algpseudocode}
\usepackage{amsmath}    
\usepackage{amssymb}    
\usepackage{amsthm}
\usepackage{graphicx}   
\usepackage{verbatim}   
\usepackage{color}      
\usepackage{subfigure}  
\usepackage{hyperref}   
\usepackage[capitalise]{cleveref}   
\usepackage{appendix}
\newcommand{\argmin}{\operatornamewithlimits{argmin}}
\newcommand{\argmax}{\operatornamewithlimits{argmax}}

\DeclareMathOperator{\Tr}{Tr}

\begin{document}
\title{Identification of simple reaction coordinates from complex dynamics}
\author{Robert T. McGibbon}
\author{Brooke E. Husic}
\author{Vijay S. Pande}
\affiliation{Department of Chemistry, Stanford University, Stanford CA 94305, USA}

\begin{abstract}
Reaction coordinates are widely used throughout chemical physics to model and understand complex chemical transformations. We introduce a definition of the natural reaction coordinate, suitable for condensed phase and biomolecular systems, as a maximally predictive one-dimensional projection. We then show this criterion is uniquely satisfied by a dominant eigenfunction of an integral operator associated with the ensemble dynamics. We present a new sparse estimator for these eigenfunctions which can search through a large candidate pool of structural order parameters and build simple, interpretable approximations that employ only a small number of these order parameters. Example applications with a small molecule's rotational dynamics and simulations of protein conformational change and folding show that this approach can filter through statistical noise to identify simple reaction coordinates from complex dynamics.
\end{abstract}
\maketitle


\section{Introduction}

The reaction coordinate --- a single collective variable that quantifies progress in a chemical reaction ---  is a ubiquitous concept in chemical kinetics.\cite{eyring1935activated, KRAMERS1940284} Reaction coordinates are, for example, required for computing reaction rates using transition state theory,\cite{eyring1935activated, KRAMERS1940284, truhlar1996current} computing kinetically meaningful free energy barriers,\cite{sichun2006effective} and accelerating conformational sampling in many biomolecular simulation protocols.\cite{Bernardi2015Enhanced, laio2002escaping, kastner2011umbrella, knight2009lambda, Torrie1977187} Their most important use, however, is often in facilitating insight into chemical reaction mechanisms.\cite{steinfeld1999chemical, RevModPhys.62.251, peters2015common}


Implicit in the concept is the notion that the measurement of reaction coordinate is dynamically informative, and provides a proxy for the rate-limiting dynamical processes of the system. Reactions in soft matter and condensed phase systems, such as the folding of a protein or an enzyme-catalyzed chemical transformation take place in a high-dimensional phase space that may include many uninvolved solute and solvent coordinates. In this regime, identification of reaction coordinates is difficult.\cite{peters2015common} Physical intuition may suffice to determine these critical degrees of freedom for low-dimensional systems, such as simple bimolecular gas-phase reactions. But in more complex processes involving tens of thousands or more atoms, rough energy landscapes, and/or solvent dynamics, methods to identify the reaction coordinate that rely merely on physical intuition or trial and error can be \emph{ad hoc} and unsystematic.\cite{cho2006p, Zhou18122001, Sheinerman17021998, Gsponer14052002}




 %

We recognize that the identification of a system's reaction coordinate(s) is invaluable for physical interpretation of complex molecular systems, that researchers now have access to extremely large data sets of unbiased molecular dynamics simulations of biologically relevant macromolecules, and that the interpretation of these data is often a major bottleneck.\cite{lane2013milliseconds} We therefore aim to develop a method to use these molecular dynamics data sets to \emph{infer} accurate and interpretable reaction coordinates. Our approach builds on time-structure based independent components analysis (tICA), a special case of the more general variational approach to conformational dynamics.\cite{schwantes2013improvements, perez2013identification} But these tICA-derived reaction coordinates can be a black box; they are difficult to interpret physically because of their abstract construction as linear combinations of a large number of structural features. In contrast, our new estimator explicitly adds a sparsity consideration into the formulation to filter through statistical noise and identify simple physical reaction coordinates from complex dynamics.

The structure of this paper is as follows: First, we define the \emph{natural reaction coordinate(s)} based on a set of intuitive mathematical properties that these collective variables should satisfy. After introducing these properties, we discuss their relationship to other commonly used definitions of the reaction coordinate. Next, we prove that this definition is satisfied by the leading eigenfunctions of an integral operator governing the ensemble dynamics.\footnote{For systems that evolve under Langevin dynamics, the operator is a backward Fokker-Planck operator.\cite{coifman2008diffusion} For a discrete-time reversible Markov chain like thermostated Hamiltonian dynamics integrated with a finite-timestep integrator, the associated operator is a backward transfer operator.\cite{schutte2001transfer}} Finally, we introduce and demonstrate a practical new estimator which can approximate these reaction coordinates as extremely sparse, interpretable, regularized linear combinations of structural order parameters.


\section{Defining the natural reaction coordinate}

Although (or perhaps because) the idea of the reaction coordinate is so widely used in chemical kinetics, the community has not always agreed on its precise meaning. A number of different definitions thereof have been proposed, including the minimum energy path or intrinsic reaction coordinate (MEP),\cite{fukui1970formulation, tachibana1980novel, quapp1984analysis, yamashita1981irc} the minimum action path (MAP),\cite{olender1997yet, heymann2008geometric, eastman2001simulation, ren2004minimum, lipfert2005protein} and the committor function.\cite{Bolhuis2002TRANSITION, dellago2002transition}

In order to proceed in the face of this definitional ambiguity, we begin from first principles and propose a set of properties that a natural reaction coordinate should satisfy for any time-homogeneous, reversible, ergodic Markov process. This approach is geared towards conformational dynamics of soft matter systems, and we make none of the assumptions common in chemical kinetics about the existence of two metastable states, about the relative importance of entropic or enthalpic barriers, about low temperature, or about the number of pathways that are possible. This level of generality does come with a trade off; it makes it impossible to leverage quasi-equilibrium approximations, and our algorithms will require equilibrium sampling. The mathematical properties which we specify require that the natural reaction coordinate (a) be a dimensionality reduction that (b) is defined only by the system's dynamics, and that (c) is the maximally predictive projection about the future evolution of the system. Below, we describe and define each of these criteria in detail.
Later on, we will show how the formulation also extends naturally to multiple orthogonal reaction coordinates.

\subsection{A dimensionality reduction from $\Omega$ to $\mathbb{R}$}

A natural reaction coordinate should be a function which maps any point in the system's full phase space to a single real number. Notating the reaction coordinate as $q$, and phase space as $\Omega$,\footnote{We use the phrase `phase space' to refer to either a position, momenta phase space, or a position-only configuration space, depending on the underlying dynamics. For thermostatted Hamiltonian or Langevin dynamics, $\Omega=\mathbb{R}^{6N}$, where $N$ is the number of atoms. For overdamped Langevin dynamics, also called Brownian or Smoluchowski dynamics, $\Omega=\mathbb{R}^{3N}$. In periodic boundary conditions, the position space is some $3N$-dimensional torus, but the exact definition of $\Omega$ is not critical for our purposes.} we may specify this as

$$
q : \Omega \rightarrow \mathbb{R}.
$$

The reason for this form is that it should be well-defined to calculate how ``far along'' the reaction coordinate any conformation is, or to speak about the mean value of the reaction coordinate for some equilibrium or non-equilibrium ensemble of conformations. Reaction coordinates taking this form include geometric or physical observables which could, in principle, be as simple as the distance between two specific atoms.

On the other hand, path-based definitions of the reaction coordinate such as the MEP or MAP do not take this form. Instead of functions from $\Omega$ to $\mathbb{R}$, a path through phase space is a function from $\mathbb{R}$ to $\Omega$. These paths map an arc length to a phase space coordinate, and the value of the reaction coordinate is undefined for all conformations in $\Omega$ that are not on this path. For the minimum energy path, this issue was discussed by Natanson \emph{et al.}, \cite{natanson1991definition} who showed that while a reaction coordinate of the form $\Omega \rightarrow \mathbb{R}$ could be defined by introducing a projection operator onto the MEP, there was considerable ambiguity in the choice of projection function. This ambiguity was present even for reactive systems containing only 3 atoms without roughness, and are exacerbated in high-dimensional and condensed phase systems. This is one factor which makes the $\Omega \rightarrow \mathbb{R}$ formulation more attractive than the $\mathbb{R} \rightarrow \Omega$ formulation.

\subsection{Uniquely determined by the dynamics}
\label{sect:uniquely_dertmined}

The natural reaction coordinate should be uniquely defined by the equations of motion that govern the underlying dynamics in $\Omega$, which include the system's Hamiltonian, boundary conditions, and integration scheme. We wish to define the natural reaction coordinate in a way that does not depend on particular ``reaction'' or ``product'' conformations or subsets of phase space.

Although it may appear intuitive to define the reaction coordinate in terms of two end points or two states, this definition has a number of formal and practical drawbacks. Subdividing phase space into non-overlapping reactant and product states, $A \subset \Omega$, $B \subset \Omega$, $A \cap B = \emptyset$, is a useful device, but this is a construct imposed by the modeller, not the underlying Hamiltonian. All experimentally measurable observables, such as ensemble averages, single-molecule time series, or time-correlation functions of a spectroscopic quantity are independent of whether the modeller labels certain regions of phase space as $A$ or $B$.


For systems containing a small number of atoms, it is often relatively obvious how these states should be determined: e.g. for a bond-forming reaction, one can simply measure whether the distance between the atoms is greater than a certain cutoff. And when the states are metastable, many quantities which might formally depend on the exact specification of the states' boundaries in fact have a very weak dependence thereon, as long as the perturbed state boundaries are still metastable.\cite{hummer2004from} But in high-dimensional systems where entropy plays a dominant role, and when confronted with significant roughness in the energy landscape on energy scales less than $k_BT$, it can be very difficult in practice to identify these metastable states. Furthermore, many systems have more than two metastable states.

Consider protein folding dynamics, where $A$ and $B$ would generally be taken to be the protein's folded and unfolded states. A number of practical definitions of the folded or unfolded state, based on metrics including root-mean-square deviations to a crystal structure, numbers of native contacts, or radii of gyration are defensible. None, however, are obviously mandated. If the definition of the natural reaction coordinate depends on the exact line-drawing between folded and unfolded, each definition of the state boundaries would lead to a slightly different natural reaction coordinate, with no criteria to judge which is optimal.

In our view, a formal defintion of the natural reaction coordinate should be unique and independent of any partitioning of phase space into regions, and only a function of the system's underlying dynamics. As a dimensionality reduction, the natural reaction coordinate should teach us about the system's metastable states, not the other way around.

\subsection{Maximally predictive projection}
\label{subsect:Maximally_predictive}

\begin{figure*}
\centering \includegraphics[width=0.8\textwidth]{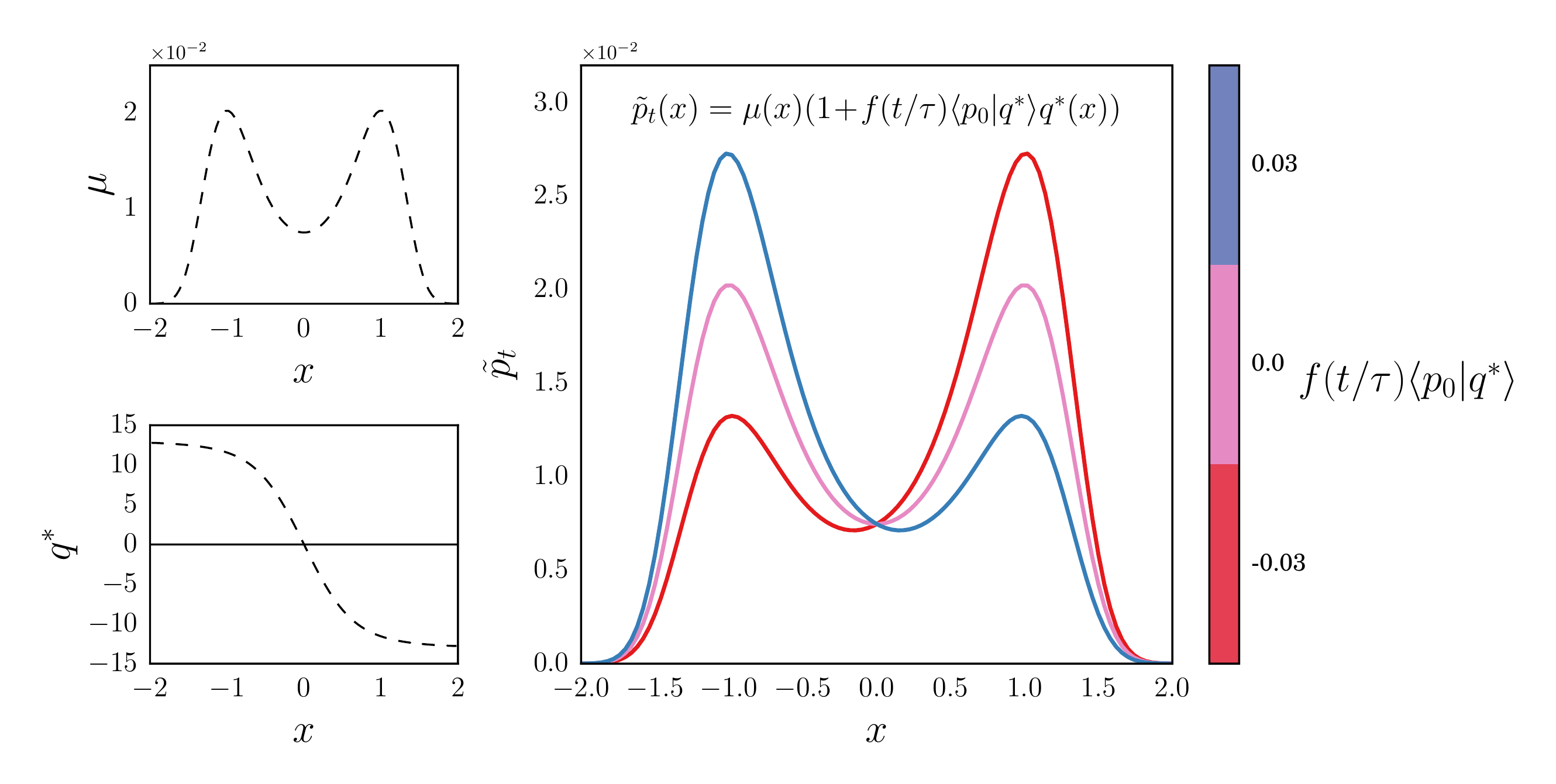}
    \caption{\label{fig:eq1} Predictions, $\tilde{p}_t$, made by the natural reaction coordinate, $q^*$, for Smoluchowski diffusion on two-well potential, $U(x) = (x-1)^2(x+1)^2$ with a uniform diffusion constant, $D=1$. The upper left panel shows the stationary distribution, $\mu(x)$, and the lower left panel shows the natural reaction coordinate, $q^*$, which changes sign between the two metastable states. The main panel shows the family of possible predictions, $\tilde{p}_t$ that can be made by \cref{eq:functionalform} using this choice of $q$, indicating the variable partitioning of density between the two basins. For an arbitrary initial distribution, $p_0$, this coordinate minimizes the worst-case predictive error about the future ensemble $p_t$ given only knowledge of the current ensemble's projection onto $q$. As discussed in \cref{sect:natureal_reaction}, $q^*$ was calculated from the second eigenfunction of the Smoluchowski operator, which was determined in this case using the FiPy PDE solver.\cite{FiPy:2009}}
\end{figure*}

Finally, the key property that we use to define the natural reaction coordinate relates to its ability to optimally predict the dynamics. Of all possible one-dimensional measurements of the state of some high-dimensional dynamical system, the natural reaction coordinate should be the most informative about the future evolution of the system. This relates to the expectation, common in chemical kinetics, that the dynamics along the reaction coordinate are rate-limiting, and that all other degrees of freedom in the system equilibrate more rapidly. The maximally predictive single coordinate will measure progress with respect to the rate-limiting bottlenecks, as the orthogonal coordinates can more reliably be assumed to be at, or near, equilibrium.

We now formalize this notion mathematically. To begin, we define the following quantities:
\begin{itemize}
\item The system has a unique equilibrium distribution over phase space, $\mu(x) : \Omega \rightarrow \mathbb{R}$. Note that $\forall x,\; \mu(x) > 0$ and $\int_\Omega dx \; \mu(x) = 1$.
\item Initially, the state of an ensemble is described by a (generally non-equilibrium) probability distribution over phase space, $p_0(x) : \Omega \rightarrow \mathbb{R}$.
\item We consider an \emph{ansatz} reaction coordinate, $q(x) : \Omega \rightarrow \mathbb{R}$, and an associated scalar, $\tau$, which will be interpreted as a timescale of the dynamics along the \emph{ansatz} reaction coordinate.
\item The scalar projection of the initial distribution, $p_0$, along the reaction coordinate is measured as $\langle q | p_0 \rangle = \int_\Omega dx\; q(x) p_0(x)$.
\item At some later time, $t>0$, the system will have evolved from $p_0$ to a new distribution over phase space, $p_t(x) : \Omega \rightarrow \mathbb{R}$, according to the underlying equations of motion for the dynamics. Note that while $p_t$ is a probability distribution, it is not a random variable; it is produced deterministically from $p_0$ and the system's equations of motion.
\end{itemize}

Now, consider the task of constructing an approximation to $p_t$. This approximation, $\tilde{p}_t$, is constrained to depend only on $\mu(x)$, $\tau$, $t$, $\langle q | p_0 \rangle$, $q(x)$, and the equilibrium mean and variance of $q(x)$. That is, given knowledge of the equilibrium distribution, the \emph{ansatz} reaction coordinate, its timescale, and \emph{no other} information about the current ensemble, $p_0$, beyond its projection onto the \emph{ansatz} reaction coordinate, our goal is to construct a prediction of the future ensemble at some later time $t$.

A basic dimensional analysis argument and the constraint that $\int_\Omega \tilde{p}_t=1$ is sufficient to establish that, assuming that $q$ is measured in a system of units such that it has mean zero in the equilibrium ensemble, the functional form of $\tilde{p}_t$ given $q$ must be
\begin{align}
\label{eq:functionalform}
\tilde{p}_t(x) = \mu(x) + f(t/\tau) \frac{\langle q| p_0 \rangle (q - \langle \mu| q \rangle)}{\langle \mu | q^2 \rangle} \mu(x),
\end{align}
where $f$ is some non-random function that is independent of $x$ and $\langle q^2| p_0 \rangle$ is the variance of $q$.
Later on, we will show that $\tilde{p}_t(x)$ is necessarily an exponential, $f(t/\tau)=e^{-t/\tau}$.
For diffusion on a double well potential, a diagrammatic example of the family of predictions, $\tilde{p}_t$, that can be made given a particular choice of $q$ is shown in \cref{fig:eq1}.

Even with full knowledge of the Hamiltonian and equations of motion, this prediction will not be exact because the one-dimensional measurement, $\langle q | p_0 \rangle$, gives incomplete information about $p_0(x)$. We define the error in the prediction, $E_{p_0}[q]$, as the $\mu^{-1}$-weighted mean squared error,
\begin{align}
E_{p_0}[q] &= ||p_t(x) - \tilde{p}_t(x)||_{\mu^{-1}}^2 \nonumber \\
&= \int_\Omega dx \; \mu^{-1}(x) (p_t(x) - \tilde{p}_t(x))^2. \label{eq:e_q0_q}
\end{align}

Note that this error depends on the arbitrary initial distribution. To remove this dependency, we consider the worst-case error by maximizing over all possible initial distributions,
\begin{align}
E[q] &= \max_{p_0} E_{p_0}[q], \label{eq:maxprobdens}\\
q^* &= \argmin_q E[q]. \label{eq:defn}
\end{align}

The functional $E[q]$ thus measures how well the measurement of an arbitrary collective variable can be used to predict the future state of the system. We define the \emph{natural reaction coordinate}, $q^*$, as the minimizer of $E[q]$. It is, in this sense, the collective variable which is maximally informative about the system's dynamics.

\subsection{Alternative Definitions}

The approach we have taken is not the only one possible. Note first the choice of error functional, \cref{eq:e_q0_q}. While it may not be initially intuitive, the $\mu^{-1}$-weighting on the norm is the logical choice for a mean squared error. It is the $\mu^{-1}$ measure, combined with detailed balance, that ensures, for example, that minimizer, $q^*$, is strictly independent of $t$ (see \cref{subsect:analysis_of_the_error_function}). A different choice, like the Kullback-Leibler divergence of Wasserstein distance would be possible,\cite{gibbs2002choosing} but lead to substantially different results. Additionally, observe that in contrast to many other formulations,\cite{berezhkovskii2005one, rhee2005onedimensional, berezhkovskii2013fiffusion} our approach is not based on the explicit construction of a one-dimensional Smoluchowski-like diffusion along the reaction coordinate.


Next, we turn our discussion to an alternative reaction coordinate definition, the committor function. This quantity was first introduced by Onsager as the splitting probability for ion-pair recombination.\cite{PhysRev.54.554} The committor is defined based on the prior identification of two non-overlapping states, $A \subset \Omega$, $B \subset \Omega$, $A \cap B = \emptyset$, which do not fully partition phase space, $A \cup B \subset \Omega$. Then, the committor, $p_A(x)$, is defined as the probability that a trajectory initiated from $x$ would enter the set $A$ before entering $B$.\cite{Bolhuis2002TRANSITION, dellago2002transition} In the context of protein folding, where $A$ is taken to be the protein's folded state, the committor is often referred to as $p$-fold.\cite{Du1998transition, pande199pathways} The committor, $p_A(x)$, takes a value of 1 for conformations inside $A$, and 0 for conformations inside $B$. The condition $\{x : p_A(x)=1/2\}$ defines a transition state ensemble or separatrix --- the set of conformations equally likely to commit to either state $A$ or state $B$.

Using the concept of the ensemble of \emph{transition paths}, which are defined as trajectory segments following the moment at which the system has exited the set $A$ and up until the systems enters the set $B$, without re-entering $A$, Hummer proved an important result.\cite{hummer2004from} He showed that, for diffusive dynamics, the probability of being on a transition path given that the system is at $x$, $\mathbb{P}(\mathrm{TP}_{AB}|x)$, is determined by the committor alone, $\mathbb{P}(\mathrm{TP}_{AB}|x) = 2p_A(x)(1-p_A(x))$. This implies also that the separatrix can be identified as the set of conformations which are most likely to be on reaction paths.

A number of computational methods build approximations to the transition path ensemble, committor or isocommittor surfaces. These include transition path sampling (TPS),\cite{dellago2002transition, Bolhuis2002TRANSITION} transition interface sampling,\cite{vanErp2005157}  and the finite temperature string method.\cite{weinan2005transition, e2005finite}

Most of the existing algorithms that identify physical reaction coordinates from molecular simulations are based on committor analysis or TPS.\cite{best2005reaction, ma2005automatic, peters2006obtaining, peters2007extensions, borrero2007reaction, peters2012inertial, peters2013reaction, peters2010peak} In the simplest version, one initializes a large collection of trajectories from isosurfaces of an \emph{ansatz} reaction coordinate and measures which of the two basins, $A$ or $B$, they commit to. If this coordinate is a good approximation to the committor, the measured splitting fraction will be narrowly peaked around the characteristic value.\cite{peters2006using} Criteria based on this observation can then be used to screen an \emph{ansatz} reaction coordinate, or optimize the parameters of a model for the reaction coordinate.\cite{ma2005automatic} More efficient maximum likelihood method which fit a parametric model for the reaction coordinate from TPS data further refine this approach.\cite{peters2007extensions, peters2012inertial}

By design, these algorithms rest on the pre-identification of the $A$ and $B$ states are not naturally suited to systems with more than two metastable states, although multiple-state extensions are available.\cite{rogal2008multiple} 
When these two states are both known \textit{a priori} and metastable, then we expect, but have not proven, that the committor function and the natural reaction coordinate are nearly equivalent. Algorithms that leverage this \textit{a priori} knowledge have the advantage of requiring significantly less sampling to converge their reaction coordinate estimators.
However, for the reasons discussed in \cref{sect:uniquely_dertmined}, we dispute the claim that the committor should be taken as the \emph{perfect} or \emph{exact} reaction coordinate.\cite{grunwald2009transition, ma2005automatic, peters2006obtaining, krivov2011numerical} The authors' experiences with large-scale simulations of protein folding and activation on Folding@Home have shown that it can be difficult to locate and precisely define these metastable states. This suggests that, for an important class of problems, the metastable states should be constructed from the output of some model, as opposed to being treated as a modelling input.\cite{shirts2000screen, voelz2010molecular, shukla2014activation} These considerations motivate our formulation of the natural reaction coordinate in a manner independent of the choice to label certain regions of phase space as $A$ or $B$.

We note that others have also defined a reaction coordinate consistent with the intuitive mathematical properties specified in Section~\ref{subsect:Maximally_predictive}, such as the subset of leading eigenfunctions estimated by diffusion maps.\cite{nadler2006diffusion, rohrdanz2011determination, Boninsegna_JCTC15} In this formulation, a map is created from sampled points in phase space by utilizing a geometric distance metric, where points that are close geometrically are expected to correspond to kinetically similar conformations. The diffusion map formulation offers the same major advantages as the natural reaction coordinate, namely that it is a dimensionality reduction that does not require any information about the system beyond its dynamics, such as knowledge of metastable states. However, it is noteworthy that results ascertained from diffusion maps are invariant to the time-indexing of trajectory frames; in other words, the duration of the path between any two conformations does not inform the analysis. As a result, diffusion maps do not provide a straightforward way to estimate the timescale for a given eigenvector. In contrast, the natural reaction coordinate defined in this work yields a mathematical relationship between a dynamical process's eigenvector and its associated timescale, and thus directly provides kinetic information about the process.

\section{A dominant eigenfunction is the natural reaction coordinate}
\label{sect:natureal_reaction}
In this section, we demonstrate that the natural reaction coordinate, as defined by the minimizer of \cref{eq:defn}, is the second leading eigenfunction of an integral operator associated with a system's Markovian dynamics in $\Omega$. For simplicity, we work here with a discrete-time Markov chain, $\{X_0, X_1, X_2, \ldots \}$, such as a typical all-atom molecular dynamics simulation with a finite time step integrator, assuming only that the prediction interval $t$ is greater than 1 step (typically on the order of 2 fs). Afterwards, we note why the same results apply if the underlying dynamics are a continuous-time Markov process, and discuss the natural generalization to multiple reaction coordinates.

\subsection{Preliminaries}
\label{subsect:theory_prelim}
The one-step dynamics of a system's Markovian evolution forward in time can be completely described in terms of a stochastic transition density kernel,
\begin{align}
    p(x, y)dy = \mathbb{P}(X_{t+1} \in B_\epsilon(y) | X_t = x),
\end{align}
where $B_\epsilon(y)$ is the open $\epsilon$-ball centered at $y$ with infinitesimal measure $dy$. Essentially, this kernel measures the conditional probability of jumping from $x$ to $y$ in one step.\footnote{For example, if the underlying dynamics are overdamped Langevin on a potential energy function $U(x)$ in units of $kT$ with unit diffusion constant, simulated using an Euler-Maruyama integrator with a unit time step, the stochastic transition density kernel, $p(x, y)dy$, would be the probability density function of a Gaussian distribution with mean $\bar{y}=x-\nabla U(x)$ and variance $\sigma^2=2$.} Integrating over the initial ensemble, $p_t$, gives a Chapman-Kolmogorov equation for the evolution of the ensemble to $p_{t+1}$,
\begin{align}
    p_{t+1}(y) &= \int_{\Omega} dx\; p_t(x) p(x, y).
\end{align}

By assumption, we consider only ergodic and reversible Markov processes. Ergodicity is the property that there do not exist two or more regions of $\Omega$ that are dynamically disconnected. That is, the integrated transition density is strictly positive, $\int_{y \in A} p(x, y) > 0$ for all $x$ and all non-empty subsets of $\Omega$, $A$. The reversibility condition is that the Markov chain obeys a detailed balance equation with respect to its stationary measure, $\mu(x)$,
\begin{align}
    \mu(x) \cdot p(x, y) = \mu(y) \cdot p(y,x).
\end{align}

For molecular dynamics, $\mu(x)$ is the equilibrium distribution associated with the thermodynamic ensemble that the system is sampling, such as the Boltzmann distribution at constant temperature, and reversibility can be interpreted as a type of generalized symmetry on the function $p(x, y)$.

The form of our maximally predictive projection formulation suggests that the reaction coordinate acts like a perturbation to the equilibrium distribution. This suggests that we consider the equations for the time evolution of a new function, $u_t(x) \equiv p_t(x)/\mu(x)$, which measures the same information as $p_t(x)$, but encoded with the excess or depletion of probability in an ensemble with respect to the stationary distribution. Applying the Chapman-Kolmogorov equation to the time evolution of $u_t$, we have
\begin{align}
    u_{t+1}(y) = \frac{1}{\mu(y)} \int_{\Omega} dx\; u_t(x) \mu(x) p(x, y).
\end{align}

This equation is taken to define the action of the one step backward transfer operator, $\mathcal{T}(1)$, which is uniquely defined by the transition density kernel,
\begin{align}
    u_{t+1}(y) = [\mathcal{T}(1) \circ u_t](y).
\end{align}

The transfer operator has many properties --- we refer the interested reader to the monograph of Sch\"utte, Huisinga, and Deuflhard for mathematical details. \cite{schutte2001transfer} For our purposes, the most relevant properties are that $\mathcal{T}(1)$ is compact and self-adjoint, and thus has a complete, countable set of real eigenfunctions and eigenvalues,
\begin{align}
    \mathcal{T}(1) \circ \psi_i = \lambda_i \psi_i,
\end{align}
which we number in decreasing order by eigenvalue magnitude. Each $\psi_i$ can be assumed to be normalized such that they are orthonormal with respect to the $\mu$-weighted inner product,
\begin{align}
    \langle \psi_i | \psi_j \rangle_\mu = \int_x dx\; \mu(x) \psi_i(x) \psi_j(x) = \delta_{ij}.
\end{align}

Furthermore, the largest eigenvalue is $\lambda_1=1$, with associate eigenfunction $\psi_1(x) = 1$, and the absolute values of the remaining eigenvalues lie within the unit interval, $|\lambda_i| < 1$.\cite{schutte2001transfer}

These properties imply that the action of $\mathcal{T}(1)$ on $u_t$ can be written as a spectral decomposition,
\begin{align}
    \label{eq:spectralT}
[\mathcal{T}(1) \circ u_t](x) = \sum_{i=1}^\infty \lambda_i \langle u_t | \psi_i \rangle_\mu \psi_i(x).
\end{align}

By repeatedly applying the single-step $\mathcal{T}(1)$ operator, we can also build the multi-step $\mathcal{T}(t)$ operator. Because of the linearity of the operator and orthonormality of the eigenfunctions, each repeated application only pulls out another factor of the eigenvalue in the sum. The spectral decomposition of $\mathcal{T}(t)$ is thus
\begin{align}
    \label{eq:spectralT2}
[\mathcal{T}(t) \circ u_t](x) = \sum_{i=1}^\infty \lambda_i^t \langle u_t | \psi_i \rangle_\mu \psi_i(x).
\end{align}

\subsection{The error functional}
\label{subsect:analysis_of_the_error_function}

\begin{figure*}
    \centering
    \includegraphics[width=0.4\textwidth]{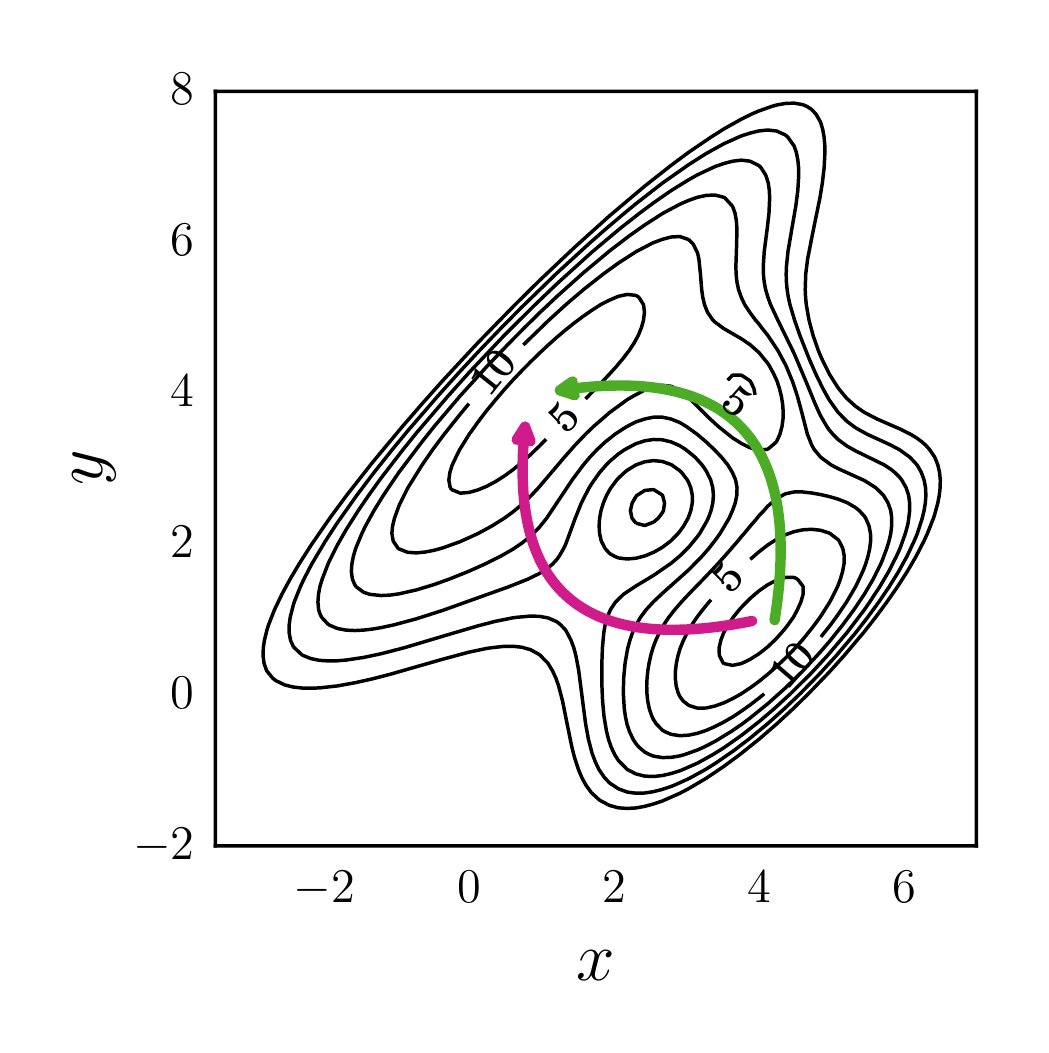}
    \includegraphics[width=0.4\textwidth]{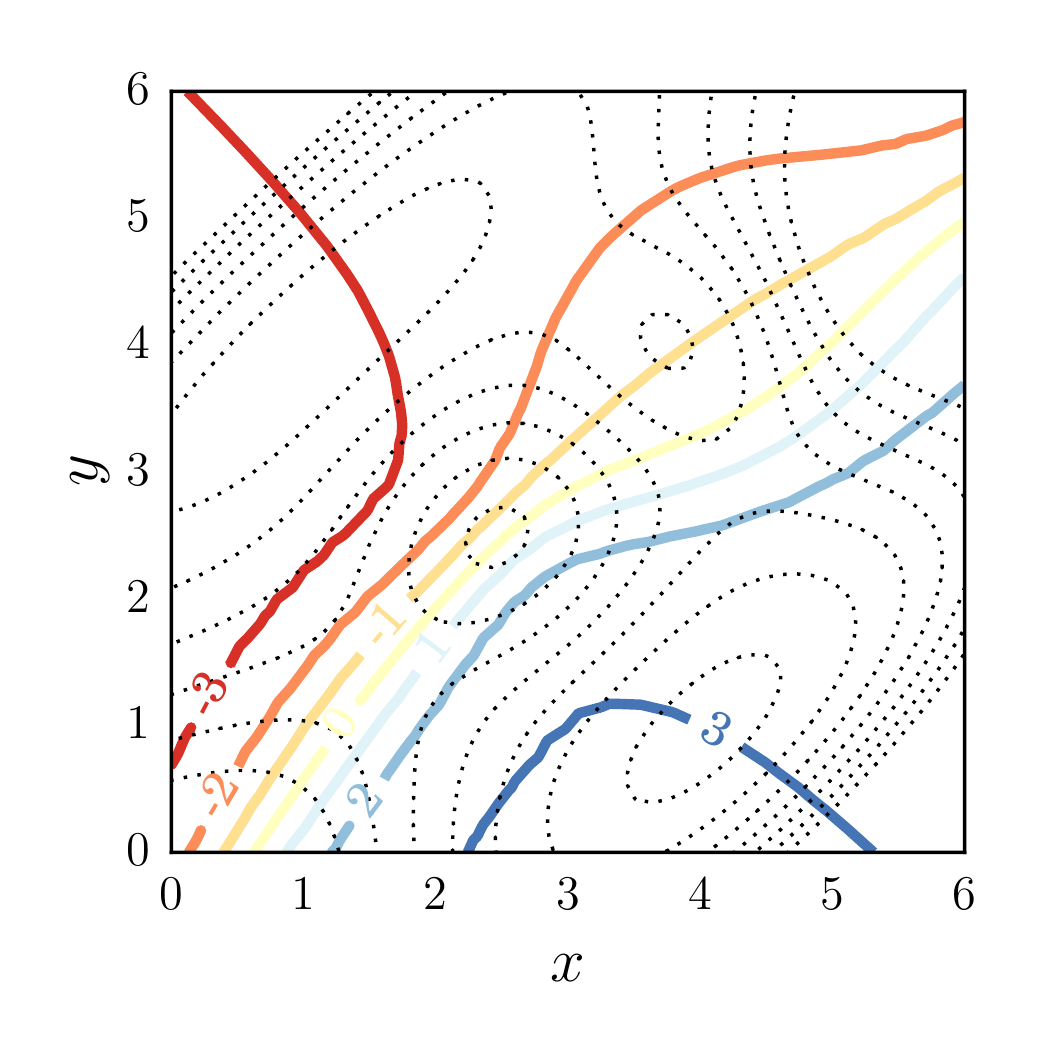}
    \caption{\label{fig:2dexample}
An example two-dimensional potential energy surface (left panel) with two of the possible pathways shown in magenta and green. The right panel shows a contour plot of the natural reaction coordinate, $\psi_2(x,y)$, for Smoluchowski dynamics at $kT=5$ with a homogeneous diffusion constant, $D=1$, overlaid on the potential energy surface, which is shown with dotted contours. We emphasize that while the natural reaction coordinate, $\psi_2 : \Omega \rightarrow \mathbb{R}$, provides a measure of progress with respect to any path between the two minima, it cannot be viewed as a single pathway itself.}
\end{figure*}

\label{subsect:Analysis_of_the_error_functional}
We now apply this spectral decomposition of the transfer operator to the analysis of the error functional from \cref{subsect:Maximally_predictive} and show that the natural reaction coordinate is equal to the second transfer operator eigenfunction, $q^* = \psi_2$.


First, observe that the form of the prediction about the future state of the system made using the reaction coordinate, \cref{eq:functionalform}, can also be written as some operator that maps $p_0 \rightarrow \tilde{p}_t$, or equivalently in $u$-notation as an approximate transfer operator, $\tilde{\mathcal{T}}(t)$, that maps $u_0 \rightarrow  \tilde{u}_t$, where $\tilde{u}_t(x) \equiv \tilde{p}_t(x)/\mu(x)$.
\begin{align}
    \tilde{u}_t &= \langle u_0 | 1 \rangle_\mu + f(t/\tau) \langle u_0 | q \rangle_\mu q(x) \label{eq:predict_u}\\
    &= \tilde{\mathcal{T}}(t) \circ u_0.
\end{align}

\noindent{}The approximate transfer operator, $\tilde{\mathcal{T}}(t)$, is rank 2; it has two non-zero eigenvalues, $1$ and $f(t/\tau)$, with associated eigenfunctions $1$ and $q(x)$ respectively.

Next, we rewrite the error functional, \cref{eq:e_q0_q} in $u$-notation as well,
\begin{align}
E_{u_0}[q] &=  \int_\Omega dx\; \mu(x) (u_t(x) - \tilde{u}_t(x))^2 \\
&= || (\mathcal{T}(t) - \tilde{\mathcal{T}}(t)) \circ u_0 ||_\mu^2 \\
\label{eq:error_functional}
E[q] &= \max_{u_0} || (\mathcal{T}(t) - \tilde{\mathcal{T}}(t)) \circ u_0 ||_\mu^2,
\end{align}
where the maximum is understood to be taken over properly normalized $u_0$, $||u_0||_\mu = 1$ instead of over probability densities as in \cref{eq:maxprobdens}. 

It follows, and is proven in Appendix~\ref{appendix:error}, that $\psi_2 = \min_q E[q]$ and that $f(t/\tau)$ can be written as $f(t/\tau) = e^{-t/\tau}$. Because it is the minimizer of $E[q]$, $\psi_2$ is the natural reaction coordinate.

\subsection{Continuous-time Markov processes}

When the generating process is a continuous-time Markov process, $\mathcal{T}(1)$ has an infinitesimal generator, $\mathcal{L}$,

\begin{align}
\mathcal{L} = \lim_{t\rightarrow 0} \frac{\mathcal{T}(t) - \mathcal{I}}{t}.
\end{align}

The set of eigenfunctions of $\mathcal{L}$ and $\mathcal{T}(t)$ are equivalent, so for these processes, $\psi_2$ can be defined in either manner.

\subsection{Multiple reaction coordinates}
\label{subsect:multiple_reaction_coordinates}
One attractive property of this definition of the reaction coordinate is that it generalizes naturally to multiple orthogonal reaction coordinates ordered by timescale

Recall that the maximally predictive projection criterion from \cref{subsect:Maximally_predictive} assumed that the approximation, $\tilde{p}_t$, was to be formed only from knowledge of the equilibrium distribution and the \emph{ansatz} reaction coordinate. The multiple coordinate generalization follows from modifying this criteria to assume knowledge of $\mu$ and the first $k-1$ eigenfunctions, $\mu$ and $\psi_2, \ldots, \psi_{k-1}$. Additionally, assume that the projection of the initial distribution onto each coordinate is available. Then, another application of the Eckart-Young Theorem shows that the maximally predictive remaining \emph{ansatz} coordinate is $\psi_{k}$. Multiple orthogonal natural reaction coordinates can thus be defined in a stepwise manner, and shown to be equal to the leading eigenfunctions, $\psi_2, \ldots, \psi_k$. In general, systems containing $k$ metastable states will have $k-1$ eigenfunctions whose associated eigenvalues are close to one, separated from the remaining eigenvalues by a so-called spectral gap.\cite{prinz2011markov}

It is reasonable to expect that for complex systems a subset of leading eigenfunctions will be required to interpret the underlying dynamical processes. For example, sufficiently long molecular dynamics simulations of any proline-containing protein should eventually sample the proline \textit{trans-cis} isomerization. Because of the partial double bond character and resulting high energy barrier for rotation about the X-Pro peptide bond (approximately 20 kcal/mol), this process is typically much slower than folding.\cite{Wedemeyer_Biochemistry02, Banushkina_JCP15} In this case, it would be necessary to interpret both $\psi_2$ and one or more eigenfunctions beyond $\psi_2$ to understand the folding process.

\subsection{Two-dimensional example}
In the left panel of \cref{fig:2dexample}, we show an example potential with two possible pathways between the dominant basins. The potential is given by the following expression,\cite{rhee2005onedimensional}
\begin{align}
\begin{split}
U(x,y) = &[1 - 0.5 \tanh(y - x)](x + y - 5)^2 \\
   &+ 0.2[((y - x)^2 - 9)^2 + 3(y - x)] \\
   &- 15e^{-(x-2.5)^2-(y-2.5)^2} - 20e^{-(x-4)^2-(y-4)^2}.
\end{split}
\end{align}

For Smoluchowski dynamics at $kT=5$ with a homogeneous diffusion constant, $D=1$, the natural reaction coordinate, $\psi_2(x,y)$, is shown with solid contour lines in the right panel of \cref{fig:2dexample}. Although $\psi_2$ can be calculated without explicitly notating any two regions $A$ and $B$ as the reactant or product state, it provides a natural measure of progress of any conformation or ensemble between the two dominant metastable states in the upper left and lower right regions of the potential.

\section{The tICA approximator}

Markov state models (MSMs) and time-structure based independent component analysis (tICA) are two widely used approximators for $\psi_2$ that can be parameterized directly from molecular dynamics trajectories.\cite{prinz2011markov, shukla2015markov, schwantes2013improvements, perez2013identification} Other popular estimators include diffusion maps and kernel tICA.\cite{nadler2006diffusion, coifman2008diffusion, rohrdanz2011determination, shwantes2015modeling, kim2015systematic}

In the tICA method, the goal is to find the optimal variational approximation to $\psi_2$ using a linear combination of basis functions. These basis functions are generally structural order parameters that can be evaluated easily for each snapshot in a simulation, such as the distance between certain pairs of atoms or some nonlinear transformation thereof, torsion angles between quartets of atoms, or root-mean-squared deviations to certain landmark conformations.

Assume that there are $m$ linearly-independent basis functions, where typical values of $m$ are in the hundreds to thousands. Without loss of generality, we assume that the basis functions have been mean-subtracted, so that they have zero mean in the equilibrium ensemble. We label the collection of basis functions as $\{\chi_j\}_{j=1}^m$.

Because $\mathcal{T}$ is self-adjoint, it can be shown that the true eigenfunction, $\psi_2$, satisfies a variational theorem,\cite{noe2013variational, nuske2014variational}
\begin{align}
\begin{split}
\psi_2 = \argmax_q\;\; &\langle q | \mathcal{T}(t) \circ q \rangle_\mu\\
\langle \mu | q \rangle &= 0 \\
\langle q | q \rangle_\mu &= 1.
\end{split}
\end{align}

Because inner products of the form $\langle q | \mathcal{T}(t) \circ q \rangle_\mu$ can be interpreted as the value of the autocorrelation function of a mean-zero, unit variance observable at time $t$,\cite{noe2013variational, schwantes2013improvements, nuske2014variational} we see as well that $\psi_2$, in addition to being the most predictive collective variable, as discussed above, is the most slowly decorrelating collective variable under the system's dynamics.

As in variational quantum chemistry methods, this quantity serves as a figure of merit for the optimization of a trial function. Expanding the \emph{ansatz} as $q = \sum_{i} a_i \chi_i$, the maximization is equivalent to the quadratic optimization problem
\begin{align}
    \label{eq:geneig}
\begin{split}
\mathbf{a}^* = \argmax_\mathbf{a}\;\;\; & \mathbf{a}^T \mathbf{C}(t) \mathbf{a} \\
\mathbf{a}^T \mathbf{\Sigma} \mathbf{a} &= 1.
\end{split}
\end{align}

The solution, $\mathbf{a}^*$, yielding the best approximation to $\psi_2$ in the span of the basis set is the generalized eigenvector associated with the largest generalized eigenvalue of the matrices $\mathbf{C}(t)$ and $\mathbf{\Sigma}$.\cite{parlett1980symmetric}

The symmetric matrix, $\mathbf{C}(t)$, and positive-definite matrix, $\mathbf{\Sigma}$, have elements given by,
\begin{align}
    C_{ij}(t) &= \langle \chi_i | \mathcal{T}(t) \circ \chi_j  \rangle_\mu = \mathbb{E}\left[ \chi_i(X_t) \cdot \chi_j(X_0) \right] \\
    \Sigma_{ij} &= \langle \chi_i | \chi_j \rangle_\mu = \mathbb{E}\left[ \chi_i(X_0) \cdot \chi_j(X_0)\right],
\end{align}
where the expectations are understood to be taken over the stochastic process. As discussed in detail by \citet{schwantes2013improvements} and \citet{perez2013identification}, the matrix elements can be estimated by empirical averages over the snapshots in molecular dynamics trajectories. The matrix $\mathbf{C}(t)$ is a collection of time-lagged correlations between the basis functions, and $\mathbf{\Sigma}$ is a covariance matrix of the basis functions. In \cref{appendix:covariance}, we discuss the use of shrinkage estimators in approximating $\mathbf{\Sigma}$ from timeseries data.

\section{A sparse approximator for the dominant eigenfunction}

%

The tICA method has one obvious drawback: the solution, our approximate natural reaction coordinate, is a linear combination of all $m$ basis functions, and the loadings are typically non-zero. This makes the solutions difficult to interpret in a mechanistic manner, because hundreds or thousands of different interatomic distances and/or torsion angles, for example, have been combined together into a single collective variable. Because an important property of reaction coordinates is their role in facilitating physical interpretation of the underlying molecular system, we consider it desirable to reduce the number of explicitly used variables.

These same interpretability issues arise with numerous methods in machine learning and statistics. For example, in multivariate linear regression, a response variable is modeled as the linear combination of input variables. Interpretable models, with only a small number of non-zero coefficients, can be obtained using variable selection methods such as the lasso.\cite{tibshirani1996regression}

In this section, we introduce a new \emph{sparse} approximator for $\psi_2$. The solution will share the same form as the tICA approximation, $q(x) = \sum_{i} a_i \chi_i(x)$, except that the vast majority of the expansion coefficients, $a_i$, will be zero. This method naturally extends to sparse approximators for each of the other leading eigenfunctions, $\psi_3, \ldots, \psi_k$.

One general approach for building sparsity-inducing estimators is to augment the objective function --- in our case, \cref{eq:geneig} --- with a regularization term that penalizes model complexity and steers the optimization towards solutions that fit the data well, but also remain simple. By scaling the strength of this term, the modeller can trade off between the two goals.

Arguably the most natural sparsity-inducing regularizer would be the $\ell_0$ norm, a penalty proportional to the number of non-zero elements in the solution vector. Unfortunately, $\ell_0$-penalized problems generally require an NP-hard combinatorial search. For many problems, such as linear regression, the most common numerically-tractable regularizers which lead to sparse solutions are based on the $\ell_1$ norm, which is sometimes interpreted as a relaxation of $\ell_0$.\cite{efron2004least, tropp2006just}

However, both the $\ell_0$ and $\ell_1$ versions of \cref{eq:geneig} are unsuitable. As discussed by \citet{sriperumbudur2011majorization}, the addition of either an $\ell_0$ or $\ell_1$ penalty to the \cref{eq:geneig} objective leads to the intractable problem of maximizing a non-concave objective function. They considered an alternative relaxation of the $\ell_0$ penalty,
\begin{align}
    ||\mathbf{x}||_0 = \sum_{i=1}^m 1_{\{|x_i| \neq 0\}} = \lim_{\epsilon \rightarrow 0} \sum_{i=1}^m \frac{\log(1+|x_i|/\epsilon)}{\log(1+1/\epsilon)}.
\end{align}

\begin{figure}
    \centering
    \includegraphics[width=0.5\textwidth]{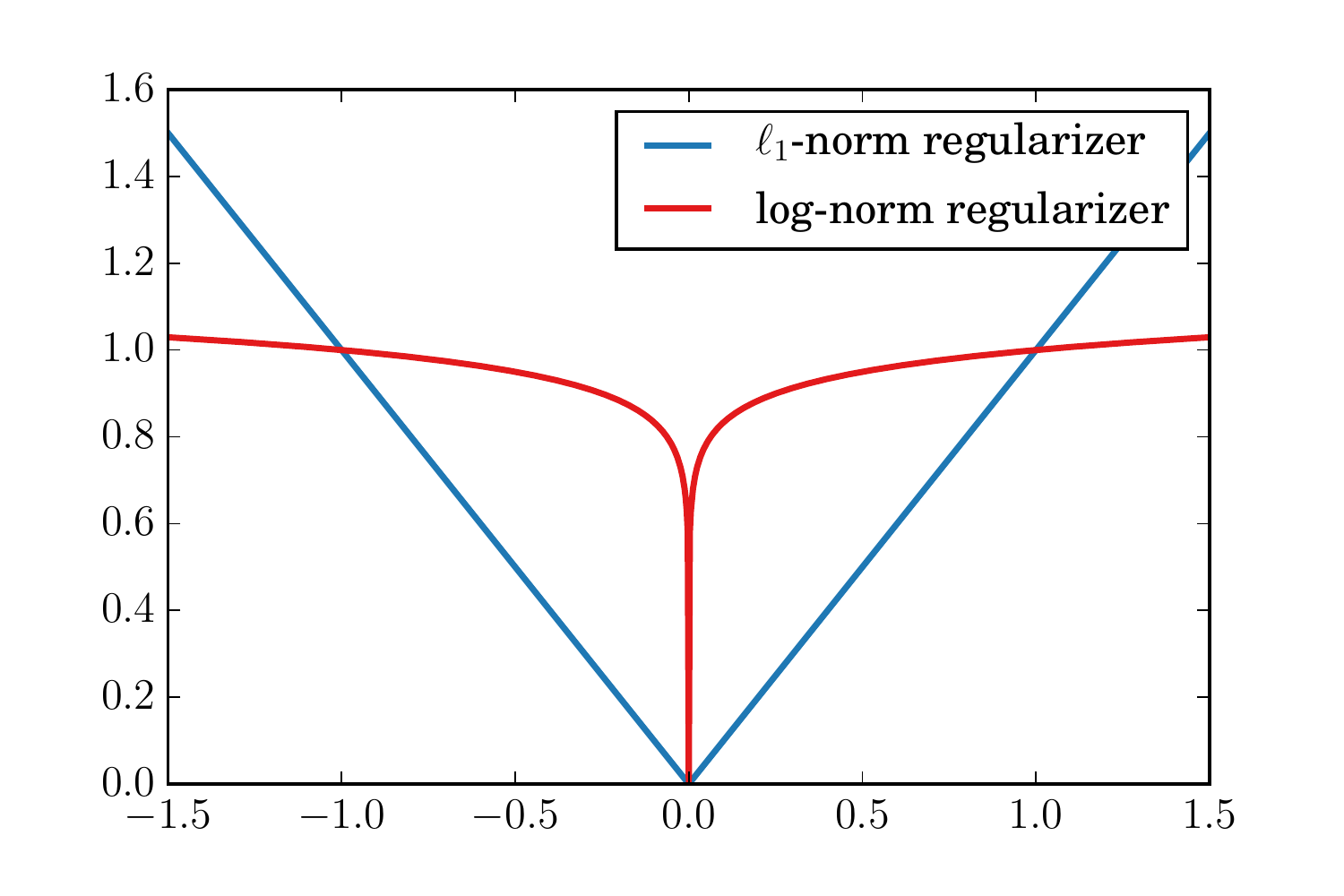}
    \caption{\label{fig:1} The log-norm regularizer used in this work, $\frac{\log(1+|x|/\epsilon)}{\log(1+1/\epsilon)}$, with $\epsilon=10^{-6}$, as compared to the $\ell_1$ norm. The log-norm is a closer approximation to the $\ell_0$ norm, and is attractive computationally for this problem because it leads to a more efficient optimization algorithm than the $\ell_1$.}
\end{figure}

Choosing a fixed $\epsilon>0$ yields a regularizer that is concave (see \cref{fig:1}), which is a property that will allow the sparse tICA method with this choice regularizer to be optimized efficiently as a difference of convex programs.\cite{horst1999dc} Therefore, to define this sparse tICA algorithm, we adopt the following formulation:\footnote{At this point, we switch the notation slightly for clarity of presentation. $\mathbf{x} \in \mathbb{R}^m$ will be the vector of sparse tICA expansion coefficients being optimized, and we take the $t$-dependence of $\mathbf{C}(t)$ to be implicit, so we simply use the notation $\mathbf{C}$.}
\begin{align}
\label{eq:sticaopt}
\begin{split}
\operatornamewithlimits{maximize}_\mathbf{x} \hspace{2em}& \mathbf{x}^T \mathbf{C} \mathbf{x} - \rho \sum_{i=1}^m \frac{\log(1+|x_i|/\epsilon)}{\log(1+\epsilon)} \\
\text{ subject to} \hspace{2em}& \mathbf{x}^T \mathbf{\Sigma} \mathbf{x} \leq 1,
\end{split}
\end{align}
where $\rho\geq0$ is the regularization strength. At $\rho=0$, the problem reduces to standard tICA. Larger values of $\rho$ will induce sparsity in the solution vectors.

Investigating sparse generalized eigenvalue problems, \citet{sriperumbudur2011majorization} showed that \cref{alg:1} is a globally convergent method for solving \cref{eq:sticaopt}. The algorithm is iterative, and refines an initial guess. Each iteration requires solving \cref{eq:qcqp}, a quadratically-constrained quadratic program (QCQP).

\begin{algorithm}[H]
\begin{algorithmic}
\Require $\mathbf{C}$ is a $n \times n$ real symmetric matrix, $\mathbf{\Sigma}$ is a $n \times n$ positive definite matrix, $\rho>0$, $\epsilon>0$
\Ensure $\mathbf{D}(\mathbf{w}^{(l)})$ be a diagonal matrix with $(w_1^{(l)},\dots,w_n^{(l)})$ as its principal diagonal, $\lambda_{min}(\mathbf{C})$ be the smallest eigenvalue of the matrix $\mathbf{C}$
\State Choose $\tau > \max(0, -\lambda_{min}(\mathbf{C}))$, $\mathbf{x}^{(0)} \in \{\mathbf{x} : \mathbf{x}^T \mathbf{\Sigma} \mathbf{x} \leq 1 \}$
\State $\rho_\epsilon = \rho / \log(1+\epsilon^{-1})$
\While{not converged} \\
\State $w_i^{(l)} \gets \rho_e\tau^{-1}  (|x_i^{(l)}| + \epsilon)^{-1}$
\State $\mathbf{b}^{(l)} \gets (\tau^{-1} \mathbf{C} + \mathbf{I}_n) \mathbf{x}^{(l)}$
\begin{align}
\;\;\;\mathbf{x}^{(l+1)} \gets &\operatornamewithlimits{argmin}_\mathbf{x} &&||\mathbf{x} - \mathbf{b}^{(l)}||_2^2 + ||\mathbf{D}(\mathbf{w}^{(l)}) \mathbf{x}||_1 \label{eq:qcqp} \\
&\text{subject to}&  &\mathbf{x}^T \mathbf{\Sigma} \mathbf{x} \leq 1 \nonumber
\end{align}
\EndWhile
\caption{\label{alg:1} \citet{sriperumbudur2011majorization}}
\end{algorithmic}
\end{algorithm}

These QCQPs are convex. When the number of basis functions, $m$, is small (less than a few hundred), we have found that they can be solved quickly and with high accuracy by off-the-shelf convex optimization libraries. However, for sparse tICA, our interest is in searching for sparse linear combinations from libraries of many thousands of possible structural order parameters. In this regime, more efficient algorithms are necessary.

\section{An ADMM solver for the QCQP subproblem}

We now derive a new, efficient solver for \cref{eq:qcqp} using the alternating direction method of multipliers (ADMM). ADMM is a general method for constructing optimization algorithms for problems of the form

\begin{equation}
\begin{aligned}
& \underset{\mathbf{x},\mathbf{z}}{\text{minimize}}
& & f(\mathbf{x}) + g(\mathbf{z}) \\
& \text{subject to}
& & \mathbf{A}\mathbf{x} - \mathbf{B}\mathbf{z} = \mathbf{c}
\end{aligned}
\end{equation}
where $f(\mathbf{x})$ and $g(\mathbf{z})$ are convex, but not necessarily smooth, functions. See \citet{boyd2011distributed} for a comprehensive review. We take $f(\mathbf{x})$ to be the original objective function from \cref{eq:qcqp},
\begin{align}
f(\mathbf{x})= \frac{1}{2}||\mathbf{x}-\mathbf{b}||_2^2 + ||\mathbf{D}(\mathbf{w})\mathbf{x}||_1,
\end{align}
where $\mathbf{D}(\mathbf{w})$ is matrix with the vector $\mathbf{w}$ along the diagonal, and $g(\mathbf{z})$ to encode the constraint,
\begin{align}
g(\mathbf{z}) = \begin{cases} 0 &\text{ if } \mathbf{z}^T\mathbf{\Sigma} \mathbf{z} \leq 1 \\ \infty &\text{ otherwise,} \end{cases}
\end{align}
where $\mathbf{A}=\mathbf{B}=\mathbf{I}_n$, and $\mathbf{c}=0$. The ADMM algorithm, in so-called scaled form, consists of the following iterations.
\begin{align}
\mathbf{x}^{(k+1)} &= \argmin_\mathbf{x} \left(f(\mathbf{x}) + \frac{\varrho}{2}||\mathbf{x} - \mathbf{z}^{(k)} + \mathbf{u}^{(k)}||^2_2 \right) \label{eq:xopt}\\
\mathbf{z}^{(k+1)} &= \argmin_\mathbf{z} \left(g(\mathbf{z}) + \frac{\varrho}{2}||\mathbf{x}^{(k+1)} - \mathbf{z} + \mathbf{u}^{(k)}||^2_2 \right) \label{eq:zopt} \\
\mathbf{u}^{(k+1)} &= \mathbf{u}^{(k)} + \mathbf{x}^{(k+1)} - \mathbf{z}^{(k+1)}, \nonumber
\end{align}
where $\varrho$ is a scalar that acts like a step size parameter, and can be adjusted over the course of the optimization to maintain stability.

By splitting the objective function into two parts, $f$ and $g$, the algorithm can alternate taking steps that minimize over the variables $\mathbf{x}$ and $\mathbf{z}$ separately, with the $\mathbf{u}$ variable serving to pull these variables towards each other and enforce the constraint that $\mathbf{x}=\mathbf{z}$ at convergence.

The advantage of this formulation is that, as we now show, both the $\mathbf{x}$ and the $\mathbf{z}$ optimization steps can be performed very efficiently.

\subsection{ADMM $\mathbf{x}$ update}
The $\mathbf{x}$ optimization, \cref{eq:xopt}, can be rewritten as
\begin{align}
\argmin_\mathbf{x} \frac{1}{2}||\mathbf{x} - \mathbf{b}||_2^2 + ||\mathbf{D}(\mathbf{w})\mathbf{x}||_1 + \frac{\varrho}{2}||\mathbf{x} - \mathbf{v}||^2_2, \label{eq:qcqp_x}
\end{align}
where $\mathbf{v}=\mathbf{z}^{(k)}-\mathbf{u}^{(k)}$. This function is component-wise separable over the elements of $\mathbf{x}$, $f(\mathbf{x}) = \sum_i f_i(x_i)$. The minimization, \cref{eq:qcqp_x},  can thus be carried out as $n$ separate scalar minimizations,
\begin{align}
\argmin_{x_i} \frac{1}{2}(x_i-b_i)^2 + w_i |x_i| + \frac{\varrho}{2}(x_i-v_i).
\end{align}

Although this objective function is not differentiable, it is a simple application of subdifferential calculus to compute a closed-form expression for the minimizer (see Ref.~\onlinecite{Rockafellar_Book70}, \textsection 23 for background). The explicit solution is
\begin{align}
x_i = \frac{1}{\varrho+1} S_{w_i}(b_i + \varrho v_i),
\end{align}
where $S$, the soft-thresholding function, is defined as
\begin{align}
S_{\kappa}(a) = \begin{cases}
a - \kappa \text{ if } a > \kappa \\
0 \text{ if } |a| \leq \kappa \\
a + \kappa \text{ if } a < -\kappa.
\end{cases}
\end{align}

This simple form and component-wise separability means that the ADMM $\mathbf{x}$ update can be computed extremely rapidly.

\subsection{ADMM $\mathbf{z}$ update}
Because $g(\mathbf{z})$ is a hard boundary function, the $\mathbf{z}$ update, \cref{eq:zopt}, can be interpreted as the projection of a point $\mathbf{a} = \mathbf{x}^{(k+1)}+\mathbf{u}^{(k+1)}$ onto the constraint set, $\{\mathbf{z} : \mathbf{z}^T\mathbf{\Sigma}\mathbf{z} \leq 1\}$, a hyper-ellipsoid. The problem  can be rewritten as
\begin{align}
\mathbf{z}^* = \begin{aligned}
&\argmin_\mathbf{z}\hspace{2em} ||\mathbf{z}-\mathbf{a}||^2 \\
&\text{subject to }\hspace{2em} \mathbf{z}^T \mathbf{\Sigma} \mathbf{z} \leq 1.
\end{aligned}
\end{align}

For the nontrivial case in which the point $\mathbf{a}$ lies outside the ellipsoid, $\mathbf{a}^T\mathbf{\Sigma}\mathbf{a} > 1$, the solution, $\mathbf{z}^*$, is on the border of the ellipsoid, ${\mathbf{z}^*}^T \mathbf{\Sigma} \mathbf{z}^* = 1$. By precomputing the eigendecomposition of $\mathbf{\Sigma}$, this can be solved efficiently using Kiseliov's method which is detailed in \cref{appendix:projection}.\cite{kiseliov1994algorithms}

An open source implementation of the estimator is available in the MSMBuilder software package at \url{http://msmbuilder.org}.

\subsection{Further orthogonal reaction coordinates}

Like tICA, our algorithm is not restricted to finding a single reaction coordinate, but can also identify sparse approximations to the other long-timescale eigenfunctions, $\psi_3, \ldots, \psi_k$. Unlike in the tICA method, in which the full set of solutions can be computed simultaneously with a single call to a standard generalized eigensolver, each sparse reaction coordinate must be estimated with a separate calculation.

As with most iterative sparse principal components analysis methods, we obtain the remaining generalized eigenvectors by subtracting the influence of the solution from the matrix $\mathbf{C}$, and then restarting optimization using the deflated matrix. The tradeoffs between methods for this deflation step have been discussed by Mackey.\cite{mackey2009deflation} Based on the recommendations therein, we have adopted Mackey's Schur complement deflation strategy.

\subsection{Hyperparameter selection and implementation notes}

In order to use sparse tICA in practice, a value of the regularization strength, $\rho$, must be chosen. When $\rho=0$, sparse tICA reduces to the standard tICA algorithm, and larger values of $\rho$ will increase the sparsity. We recommend two possible methods of choosing $\rho$. First, with cross-validation, the modeller may split the data set into two or more portions, optimize the reaction coordinate at different values of $\rho$ using one fraction of the data set, and check the value of the objective function on the left-out data set. For tICA and Markov state models, this approach was discussed McGibbon and Pande.\cite{mcgibbon2015variational} It is equally applicable to sparse tICA.

Alternatively, when the primary goal is to generate physically interpretable reaction coordinates, the modeller may choose the value of $\rho$ to bring the number of non-zero loadings down to a pre-specified number that is amenable to interpretation. When employing this strategy, we recommend that modellers watch the value of the pseudoeigenvalue (Rayleigh quotient), $\hat{\lambda}=\mathbf{x}^T\mathbf{C}\mathbf{x} / \mathbf{x}^T \mathbf{\Sigma}\mathbf{x}$. It should decrease slightly with increasing $\rho$, but dramatic drops in $\hat{\lambda}$ may indicate over-regularization.

The procedure also depends on $\epsilon>0$, which controls the shape of the regularizer. Lower values of $\epsilon$ lead to a tighter approximation of the $\ell_0$ norm, but can also lead to numerical instabilities as the derivative of the regularizer near zero goes to infinity, as can be seen in \cref{fig:1}. Empirically, we have found that $\epsilon=10^{-6}$ provides a suitable balance.

Finally, note that the scalar $\varrho$ is required during the optimization as well. This parameter affects only the convergence rate of the solver, as opposed to the final solution, and can be dynamically adjusted over the course of the optimization using standard methods described by \citet{boyd2011distributed}

\section{Examples}\label{sec:examples}
\subsection{Torsional reaction coordinate}

\begin{figure}
\centering
\includegraphics[width=0.2\textwidth]{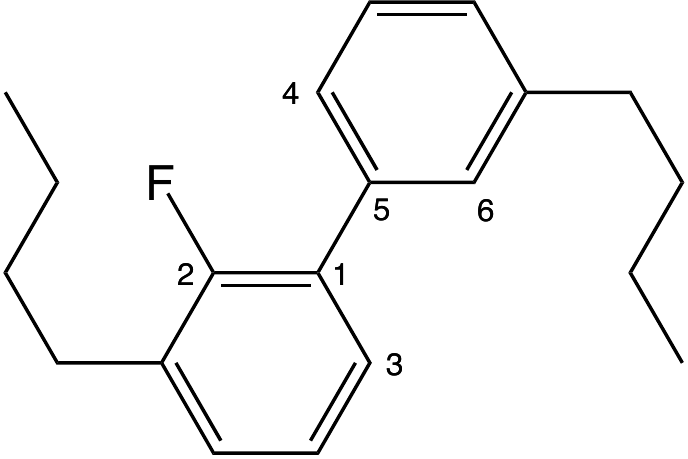}
\caption{\label{fig:biphenyl} A 2-fluorobiphenyl derivative simulated in this work. An overcomplete set of 510 internal coordinates were measured from each frame, which included four dihedral angles (described by carbons 2-1-5-4, 2-1-5-6, 3-1-5-4, and 3-1-5-6) that described the inter-ring torsion angle.}
\end{figure}

\begin{figure*}
\centering
\includegraphics[width=\textwidth]{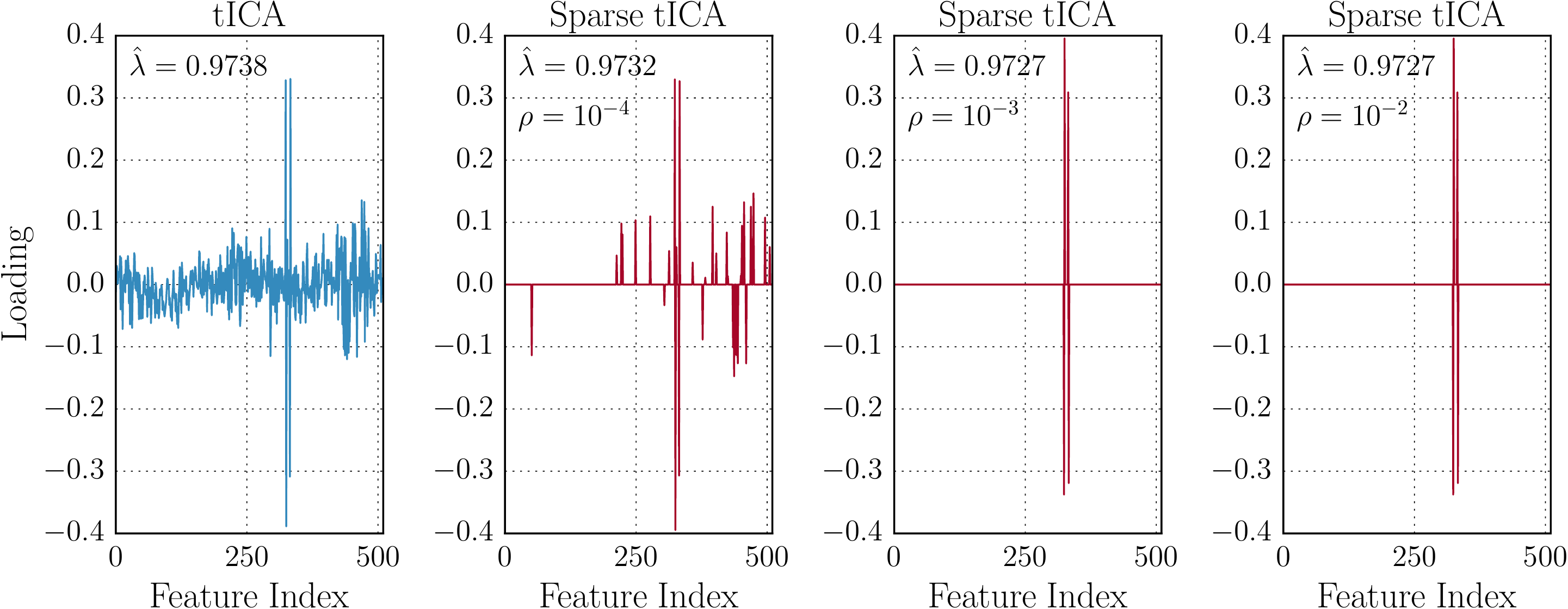}
\onecolumngrid
\caption{\label{fig:biphenyl-sparsetica} tICA and sparse tICA results for simulations of the 2-fluorobiphenyl derivative shown in \cref{fig:biphenyl} with increasing values of the regularization strength, $\rho$. The unregularized tICA results report a reaction coordinate which is a dense linear combination of all 510 input features. In contrast, with increasing values of the regularization strength, $\rho$, the sparse tICA algorithm filters out this noise to identify only the sines of the four dihedral angles that collectively characterize the inter-ring torsional reaction coordinate, with only a minor decrease in the psuedoeigenvalue, $\hat{\lambda}$.}
\end{figure*}

We demonstrate our approach on molecular dynamics simulations of a  simple 2-fluorobiphenyl derivative, shown in \cref{fig:biphenyl}. This system is interesting as a toy example because chemical intuition suggests that the rotation of the rings with respect to one another will be hindered. We anticipate the dynamics of the aliphatic tails to be faster and uncoupled to the reaction coordinate. Can our algorithm recover this sparse reaction coordinate?

After parameterization with the generalized Amber forcefield,\cite{wang2004development} we simulated the system in the gas phase for 250 ns at 290 K using a Langevin integrator with a friction coefficient of $1$ ps$^{-1}$ and timestep of 2 fs using OpenMM 6.3.\cite{eastman2013openmm} Snapshots from the simulation were saved every 20 ps. From each simulation snapshot, we recorded the values of an overcomplete set of 510 internal coordinates, which included the distances between all unique pairs of carbon atoms, measured in nanometers, the angles between pairs of bonded atoms, in radians, and the sine and cosine of the dihedral angles between all quartets of bonded atoms. After mean subtraction, these coordinates form our basis functions, $\chi_i$, for tICA and our sparse variant. Despite our chemical intuition, from an algorithmic perspective, finding the reaction coordinate for this system is something like finding a needle in a haystack.

In \cref{fig:biphenyl-sparsetica}, we show the resulting dominant eigenvector as estimated by tICA and our new approach using increasing values of the regularizer, $\rho$. The pseudoeigenvalue, $\hat{\lambda}$, is the Rayleigh quotient of the collective variable, related to its timescale by $\hat{\tau}_i = -1/ \ln \hat{\lambda}_i$. In standard tICA, this value is maximized exclusively, whereas in sparse tICA, this objective is balanced against a penalty that favors zero coefficients. We see in \cref{fig:biphenyl-sparsetica} that the tICA solution, as expected, returns a collective variable that is a linear  combination of all 510 input coordinates, with a nonzero component on each of the coordinates and significant noise.

In constrast, our sparse tICA algorithm suppresses this noise and identifies sparse collective variables that are formed from linear combinations of only a small number of the input degrees of freedom. This sparsity increases with larger values of the regularization strength, $\rho$, and only leads to a modest decrease in the approximated timescale associated with the coordinate. For $\rho=10^{-3}$ and $\rho=10^{-2}$, only four input coordinates survive. Inspection of these coordinates shows that they are the sines of the four dihedral angles that cross between the rings (atoms 2-1-5-4, 2-1-5-6, 3-1-5-4, and 3-1-5-6 in \cref{fig:biphenyl}). We interpret these results to show that sparse tICA has, without any prior chemical knowledge, filtered through a collection of structural order parameters, many of which are irrelevant in describing the slowest dynamical process of this molecule, and located the subset which can approximate the natural reaction coordinate.

\subsection{Bovine pancreatic trypsin inhibitor (BPTI)}

\begin{figure}
    \centering
    \includegraphics[width=0.5\textwidth]{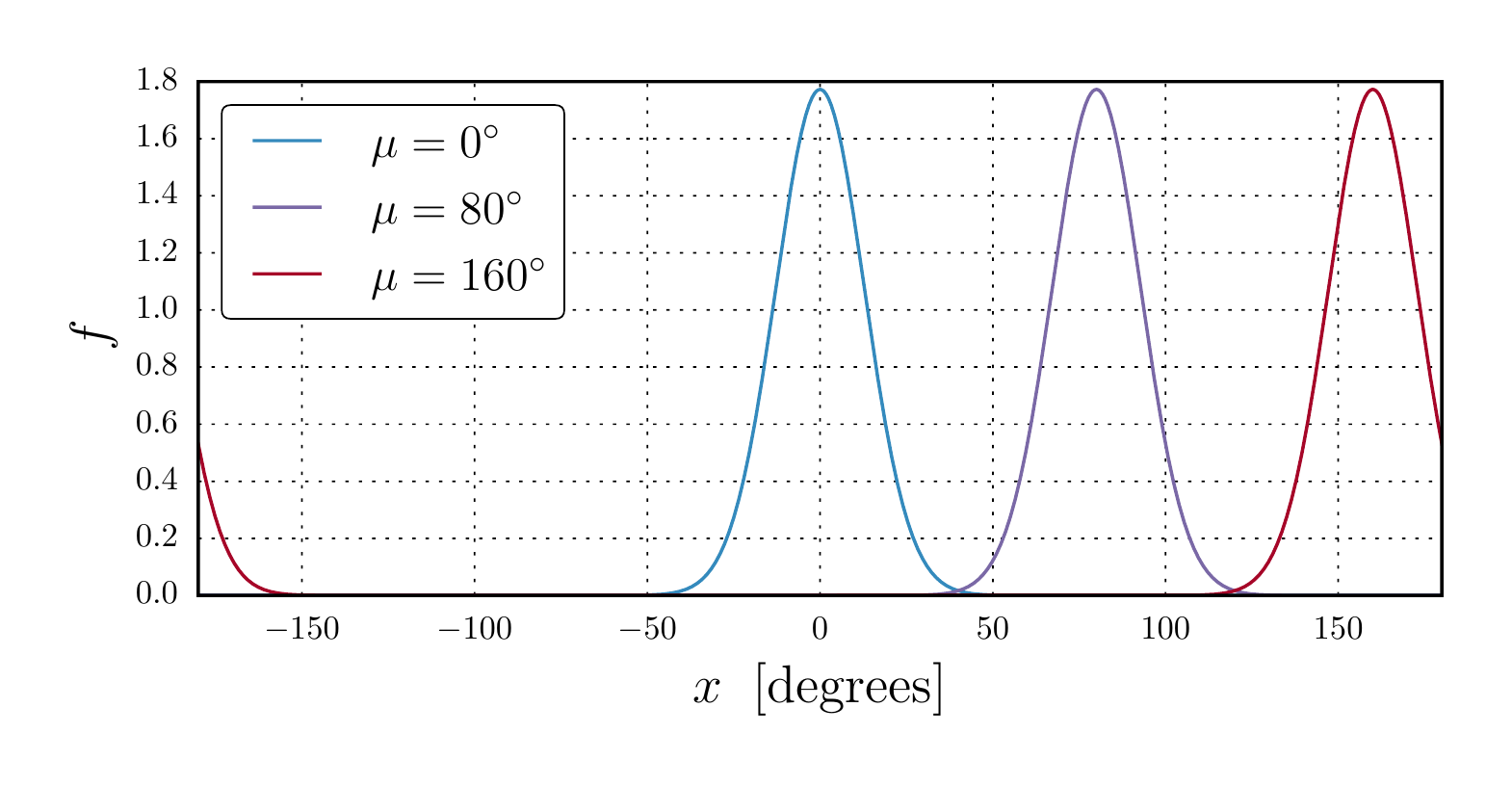}
    \caption{\label{fig:vonmises} Probability density function of the von Mises distribution with $\kappa=20$ and different values of the location parameter, $\mu$. For an angle $x$, the function is given by $f(x;\kappa) = \frac{e^{\kappa \cos(x-\mu)}}{2\pi I_0(\kappa)}$, where $I_0(\kappa)$ is the modified Bessel function of order 0. The function has a full-width at half maximum of approximately $30^\circ$.}
\end{figure}

\begin{figure}
    \includegraphics[width=0.5\textwidth]{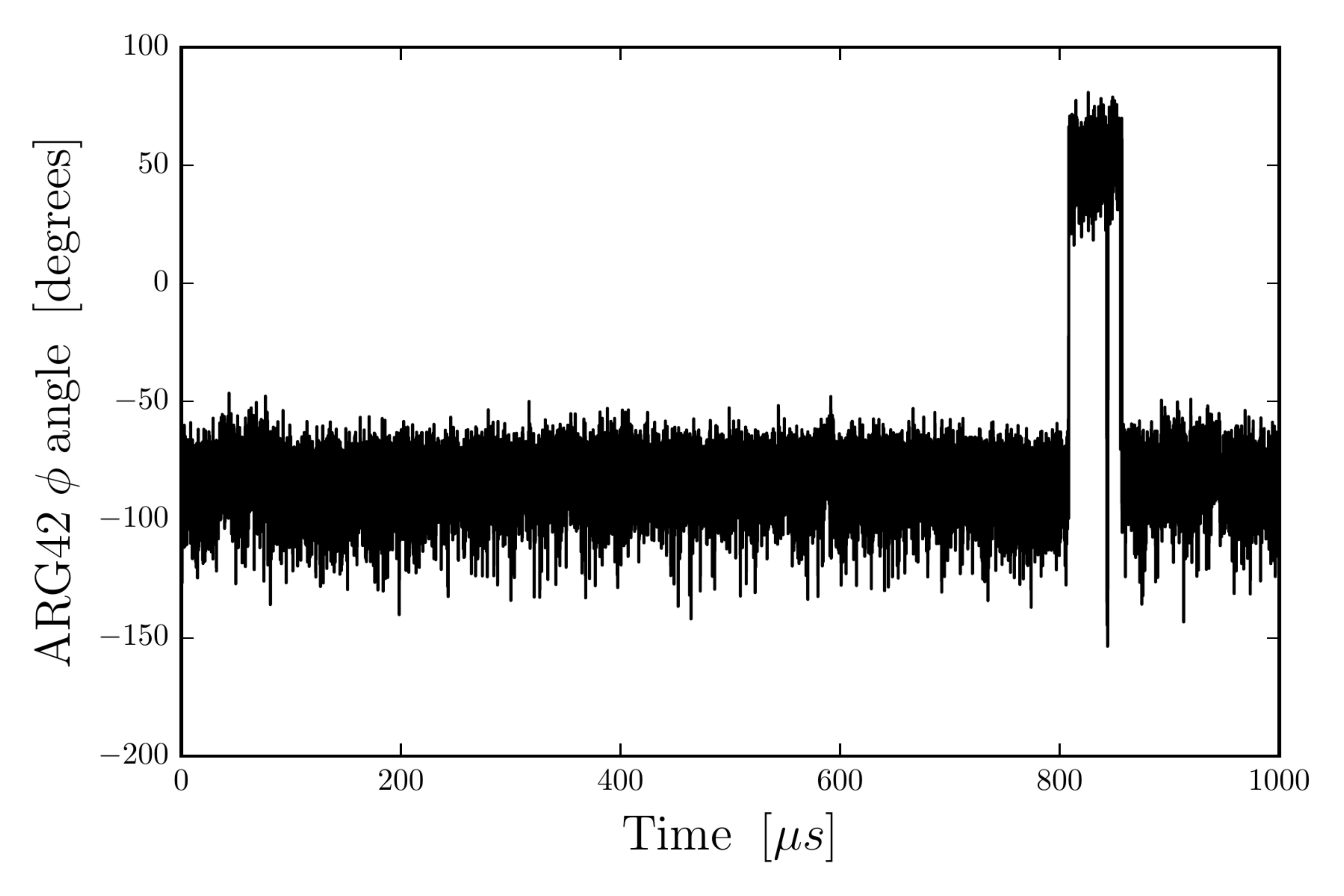}
    \caption{\label{fig:bpti_phi42_angle_timeseries} The ARG 42 $\phi$ angle over the course of the 1 ms simulation of native state dynamics of BPTI performed by D.E. Shaw Research.\cite{shaw2010atomic}. Our sparse tICA identifies this as the reaction coordinate for a process that involves the opening and hydration of the protein's core.}
\end{figure}

\begin{figure}
    \centering
    \includegraphics[width=0.4\textwidth]{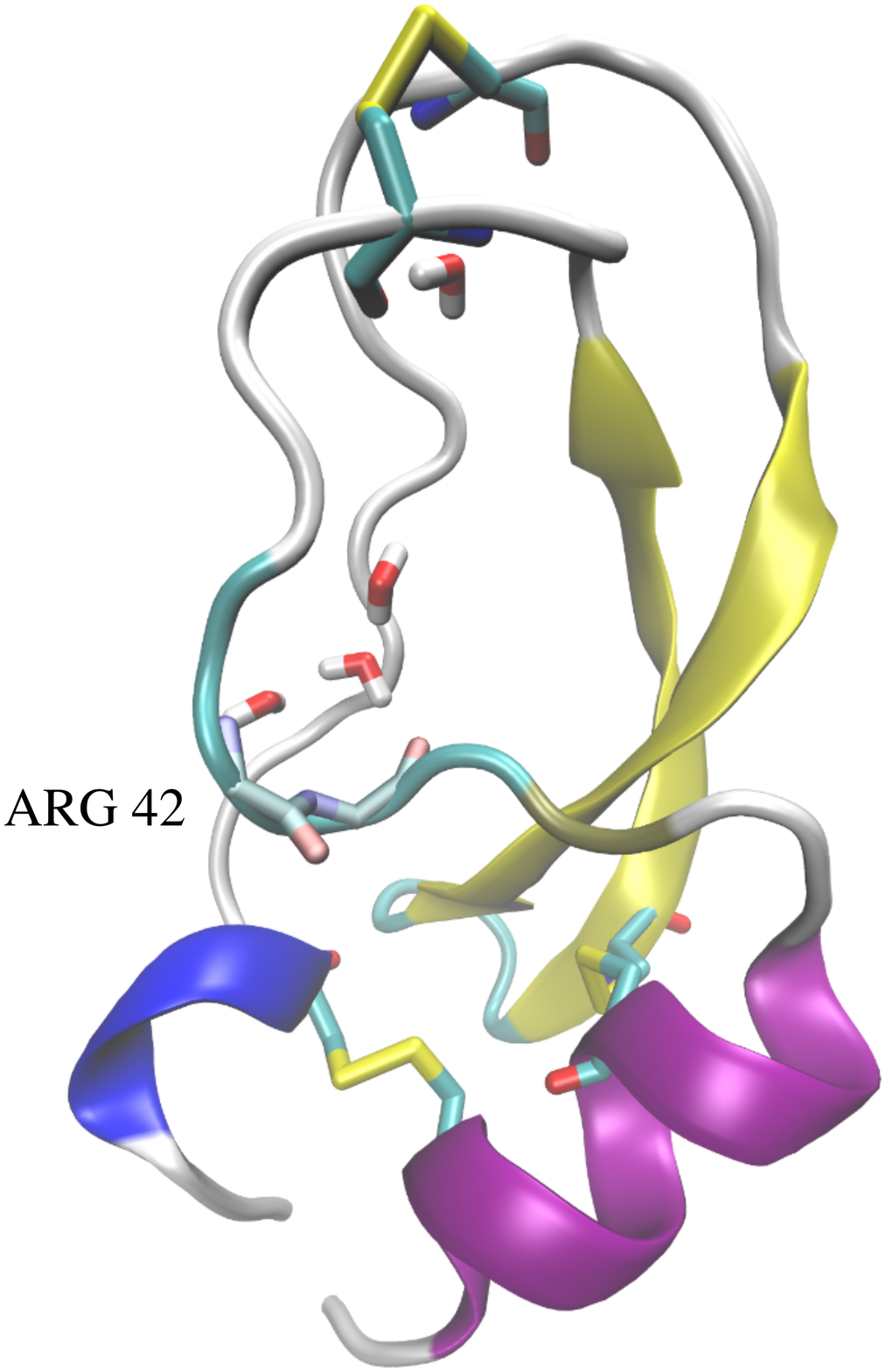} \\
    \includegraphics[width=0.4\textwidth]{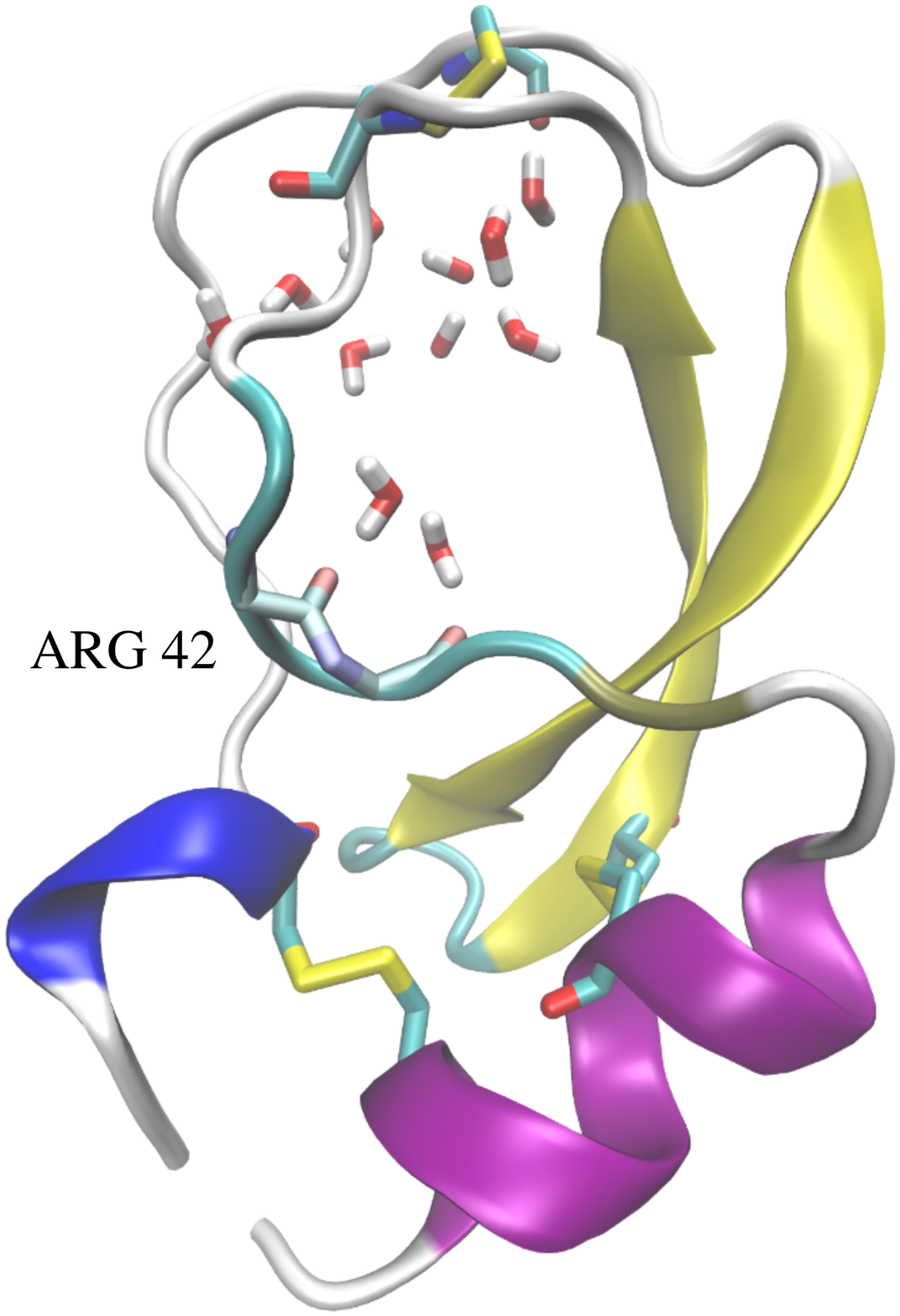}
    \caption{\label{fig:bpti_conformations} The near-native (above) and ARG42-flipped (below) conformations of BPTI from the simulation trajectory. The first panel shows the near-native conformation sampled by the majority of the simulation with a ARG42 $\phi$ angle between $-50^\circ$ and $-150^\circ$, with the expected four crystallographic waters. Nearly $800$ $\mu s$ into the simulation, the trajectory samples an alternate state in which the protein's core opens and hydrates and the crystallographic waters can exchange with the bulk. In this state, the ARG42 $\phi$ angle has flipped, putting its oxygen pointing into the now hydrated core.}
\end{figure}

In this section, we apply the sparse tICA method to analyze the native state dynamics of the bovine pancreatic trypsin inhibitor (BPTI), a small 58-residue globular protein that has been extensively investigated by experimental and computational methods. We reanalyzed the one millisecond all-atom molecular dynamics simulation performed by D.E. Shaw Research at 300K with explicit solvent.\cite{shaw2010atomic}  With its rigid disulfide bonds, the system remains folded over the course of the simulation, but samples a number of near-native states.

For each frame in the trajectory data set, sampled every 25 ns, we computed the value of an extensive set of 2880 structural order parameters from the backbone and side chain dihedral angles. For each of the 57 protein backbone $\phi$ and $\psi$ torsion angles, as well as the 46 $\chi_1$ torsion angles, we computed 18 order parameters by evaluating the probability density function of the von Mises distribution at different values of its location parameter, evenly spaced around the unit circle at $20^\circ$ increments. A subset of these functions is shown in \cref{fig:vonmises}. These functions act like softened indicator functions that wrap appropriately on $(-180^\circ, 180^\circ)$. We hypothesized that this would be a suitable basis in which to expand the reaction coordinates for BPTI, because it is well suited for expressing a function representing flux between two regions on a Ramachandran plot. Each structural order parameter in our input basis set can thus be interpreted as roughly indicating whether a particular torsion angle is within one of 18 different $\sim 30^\circ$ windows.

Using these input features, we fit a sparse tICA model with $\rho=0.005$ and observed a surprising result. The first solution depends only on the $\phi$ dihedral angle of ARG 42. The timeseries of this angle over the course of the simulation is shown in \cref{fig:bpti_phi42_angle_timeseries}, and we see that this degree of freedom makes a single dramatic flip over the course of the simulation. When we inspected conformations from this flipped state, we observed that the protein's core had opened and hydrated. While this large-scale structural change is obvious from visual inspection of the trajectory, fact that the $\phi$ angle of ARG 42 acts as a switch between these two states was unexpected. While many other degrees of freedom also change between these two states, such as the orientation of the upper disulfide linkage (visible in \cref{fig:bpti_conformations}) these degrees of freedom also fluctuate within the near-native state. It is the rare inward flip of ARG 42 which we observe to draw in solvent to hydrate the protein's small core.

\subsection{Folding of a three-helix bundle}

\begin{figure*}
\centering
\includegraphics[width=0.85\textwidth]{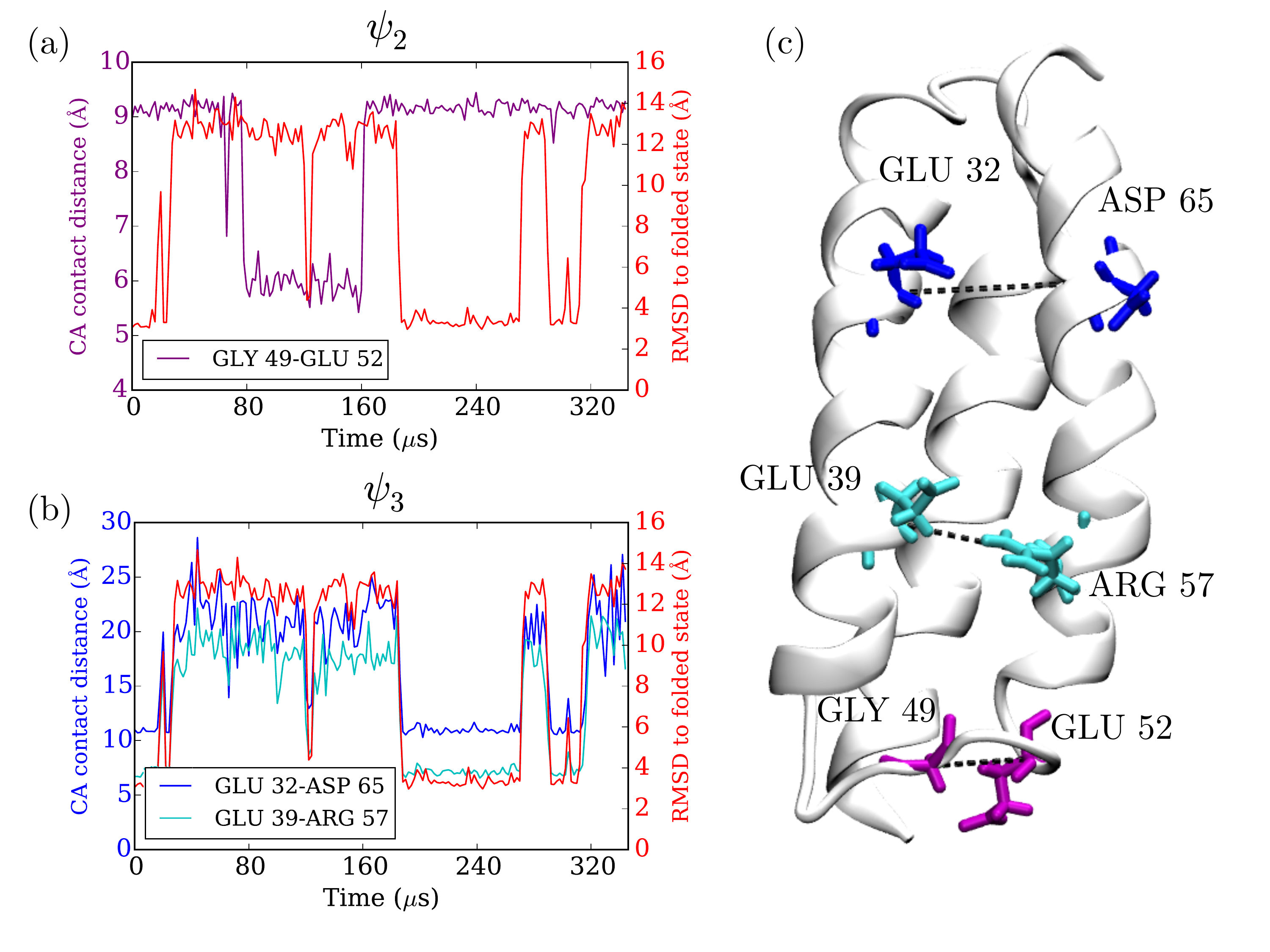}
\onecolumngrid
\caption{\label{fig:a3d}(a) Superimposing a plot of the slowest sparse tICA solution, $\psi_2$, and the RMSD of the protein conformation to the folded structure shows that the inter-residue contact distance isolated by $\psi_2$ does not correspond to the folding of $\alpha$3D. (b) Superimposing a plot of the next sparse tICA solution, $\psi_3$, and the RMSD of the protein conformation to the folded structure shows that either inter-residue contact distance implicated in $\psi_3$ serves as a suitable reaction coordinate for folding. (c) The folded state of $\alpha$3D illustrating the residue pairs defining the contact distances retained in $\psi_2$ (violet) and $\psi_3$ (blue and cyan). Only the first half of the dataset is shown.}
\end{figure*}

Next we use the sparse tICA algorithm to elucidate a specific process; in this case, the folding of $\alpha$3D, a 73-residue three-helix bundle.\cite{Walsh_PNAS99, Kubelka_COSB04} We analyzed the $\alpha$-carbon trace of a 707 $\mu$s molecular dynamics dataset for $\alpha$3D generated by \citet{LindorffLarsen_Science11}. The protein folds and unfolds 12 times over the course of the simulations. We extracted inter-residue $\alpha$-carbon distances for all pairs separated by least two residues from each frame for a total of 2485 distances. From these distances, we fit a sparse tICA model ($\rho$=0.5).

The dominant reaction coordinate, $\psi_2$, depends on just one feature: the distance between GLY 49 and GLU 52. These residues are close in the sequence and typically remain separated by about 9 $\textup{\AA}$; however, they occasionally are found within 6 $\textup{\AA}$ of each other. A plot of the GLY 49--GLU 52 distance superimposed over a plot of the conformation's root-mean-square deviation (RMSD) shows no obvious relationship between this contact distance and the folding process (Fig.~\ref{fig:a3d}a). However, trajectory events characterized by the shortening of the GLY 49--GLU 52 distance occur more rarely than folding events, and thus this contraction is the slowest process found by sparse tICA.
This slow dynamical process is intriguing but may be artifactual. Three plausible interpretations of this result are that the identified process is (1) a random artifact of unconverged sampling, (2) an artifact due to a systematic problem with the force field (as opposed to a statistical anomaly), or (3) a legitimate and newly identified slow, dynamical proess in the unfolded state of $\alpha$3d. Regardless of the correct interpretation of $\psi_2$, the algorithm identifies a new and interesting degree of freedom.

However, our intention is to use the sparse tICA algorithm to gain insight into the folding process. The second solution, $\psi_3$, isolates two residue contact pairs: GLU 32--ASP 65 and GLU 39--ARG 57. Both pairs contain residue contacts between the same two $\alpha$-helices. Fig.~\ref{fig:a3d}b shows a plot of the two contact distances comprising $\psi_3$ superimposed with the conformation's RMSD. It is clear that both the GLU 32--ASP 65 and GLU 39--ARG 57 distances serve as a sparse proxy for whether the protein is folded or unfolded. This analysis suggests that the formation of the tertiary contact between the two helices identified by $\psi_3$ is the rate-limiting step of the folding process.


\section{Conclusions}

In this work, we have introduced a defintion of the natural reaction coordinate as a function that satisfies a set of simple mathematical properties: that it (a) is a dimensionality reduction that (b) is defined only by the system's dynamics, and that (c) is the maximally predictive projection about the future evolution of the system. The definition is particularly apt for soft-matter systems in which there may be more than two metastable states, or for systems in which identifying and structurally defining the metastable states is challenging. For any time-homogeneous, reversible, ergodic Markov chain such as thermostatted molecular dynamics, these properties are uniquely satisfied by a dominant eigenfunction of the transfer operator associated with the dynamics, $\psi_2$. This eigenfunction is also the most slowly decorrelating collective variable in the system. Subsequent, orthogonal reaction coordinates for other long-timescale dynamical processes are described by the leading eigenfunctions $\psi_3$ and following.

We developed a practical new estimator that builds upon the tICA method for estimating these eigenfunctions. Like tICA, this estimator is used to post-process molecular dynamics trajectories. Unlike the variational tICA method which constructs an approximation to these eigenfunctions using a linear combination of structural order parameters in which all of the coefficients are generally non-zero, our estimator finds sparse solutions. It is thus able both to filter through inevitable statistical noise and identify simple, interpretable strutural order parameters that approximate these natural reaction coordinates, without any prior knowledge of the system.

Application of this method to molecular dynamics simulations of a 2-fluorobiphenyl derivative and BPTI show that the approach can identify reaction coordinates for the slow dynamical processes in these data sets that are readily interpretable. In BPTI, we see that opening and hydration of the protein core is controlled by a flip of a single backbone $\phi$ angle at ARG 42.

When applying sparse tICA to folding simulations of $\alpha$3D, we find that a nondominant reaction coordinate, $\psi_3$, serves as a reaction coordinate for folding while the dominant reaction coordinate, $\psi_2$, instead captures a seemingly unrelated, rare contraction of the distance between two residues close in the protein sequence. This example highlights that the desired reaction coordinate many not be the first (i.e.~slowest) solution to the algorithm. Furthermore, when the process corresponding to the dominant reaction coordinate seems unrelated to the process of interest, it may indicate that the system dynamics have been insufficently sampled, or motivate inspection of the force field parameters related to the features controlling $\psi_2$.

We anticipate that this method will be useful for the analysis of today's large molecular dynamics data sets. An implementation of this estimator is available in the MSMBuilder software package at \url{http://msmbuilder.org/} under the GNU Lesser General Public License.

\section*{Acknowledgments}
The authors thank Thomas J. Lane for helpful discussions made during the preparation of this manuscript, Ariana Peck and Carlos X. Hern\'{a}ndez for invaluable copy editing, and the National Institutes of Health under Nos. NIH R01-GM62868 for funding. We graciously acknowledge D.E. Shaw Research for providing access to the BPTI trajectory data set.

\appendix
\section{Analysis of the error functional}
\label{appendix:error}

To prove by why $\psi_2 = \min_q E[q]$, (Eqn.~\ref{eq:error_functional}), observe that for any $q$, there exists a function $v(x)$ in the span of the first three eigenfunctions of $\mathcal{T}$, $v=a_1 \psi_1+ a_2 \psi_2 + a_3 \psi_3$, which is normalized, $\langle v | v \rangle_\mu = 1$, and which is in the null space of $\tilde{\mathcal{T}}$, $\tilde{\mathcal{T}} \circ v = 0$.\footnote{To be concrete, set $a_1 = 0$, and choose $a_2$ and $a_3$ to satisfy \protect{$a_2 \langle \psi_2 | q \rangle_\mu  = - a_3 \langle \psi_3 | q \rangle_\mu$ and $a_2^2 + a_3^2 = 1$}.} Since $E[q]$ is the maximum of $E_{\mu_0}[q]$ over all possible $\mu_0$, it also must be greater than the error incurred for this particular starting distribution, $\mu_0=v$. Thus,
\begin{align}
E[q] &\geq || (\mathcal{T}(t) - \tilde{\mathcal{T}}(t)) \circ v ||_\mu^2 \\
&= || \mathcal{T}(t) \circ v ||_\mu^2 \\
&= \sum_{i=1}^3 \lambda_i^{2t} a_i^2 \\
&\geq \lambda_3^{2t},
\end{align}
where the third line only includes a sum up to $i=3$ because, by construction, $v$ is in the span of the first three eigenfunctions. The final line follows because of the ordering of the eigenvalues and the normalization of $v$, implying $\sum_{i=1}^3 a_i^2 = 1$.

Interpreting this inequality, we see that the worst-case prediction error for any \emph{ansatz} reaction coordinate, $q$, is always greater than or equal to $\lambda_3^{2t}$. Furthermore, for the particular choice $q=\psi_2$ and $f(t/\tau)=\lambda_2^t$, the equality is achieved, $E[\psi_2] = \lambda_3^{2t}$.\footnote{To demonstrate that $E[\psi_2] = \lambda_3^{2t}$, note that for this choice of $q$ and $f(t/\tau)$, $\tilde{\mathcal{T}}(t)$ is equal to the sum of the first two terms in the spectral decomposition of $\mathcal{T}(t)$. The squared spectral norm of the difference between the two operators is the square of the largest eigenvalue of the difference operator. The first two eigenpairs having been subtracted out, the square of the largest remaining eigenvalue is $\lambda_3^{2t}$.} If we define $\tau \equiv -1/\ln \lambda_2$, $f(t/\tau)$ can be written as $f(t/\tau) = e^{-t/\tau}$. Therefore $\psi_2$ is the natural reaction coordinate, the minimizer of $E[q]$.

The reader may recall that this argument is equivalent to the Eckart-Young Theorem on the optimal low-rank approximation of a matrix.\cite{eckart1936approximation} For self-adjoint linear operators, the original result is by Schmidt.\cite{schmidt1907zur} See Courant and Hilbert (pp.~161),\cite{courant2008methods} and \citet{micchelli1971some} for further details.

\section{Covariance matrix estimation}
\label{appendix:covariance}

In this section, we discuss some issues related to the estimation of the covariance matrix, $\mathbf{\Sigma}$, from timeseries data such as molecular dynamics simulations. If we consider a single trajectory of length $N$ and collect the results of the evaluation of each of the zero-meaned $m$ basis functions on each of the $T$ snapshots into a matrix, $\chi \in \mathbb{R}^{m \times N}$, the standard estimator for $\mathbf{\Sigma}$ would be the sample covariance matrix,
\begin{align}
    \mathbf{S} = \frac{1}{N-1} \chi \chi^T.
\end{align}

Covariance matrix estimation is a ubiquitous problem common to many fields of science and engineering, and a number of issues with this estimator are known. In particular, results from random matrix theory suggest that the eigenspectrum of the estimated covariance matrix, $\hat{S}$, is over-dispersed with respect to the true value. That is, its large eigenvalues are too large, and its small eigenvalues are too small. For a fixed number of basis functions, $m$, the sample eigenvalues can be shown to converge to the true eigenvalues as $N$ goes to infinity,\cite{anderson1963} but when $m$ is allowed to grow with $N$, keeping $m/N$ fixed, results such as the Mar\v{c}kenko-Pastur law suggest that the sample eigenvalues are not effective estimators, and do not converge to the true eigenvalues.\cite{johnstone2001}

In the context of a weight matrix in a generalized eigenvalue problem, misestimation of the small eigenvalues of $S$ is particularly problematic. The generalized eigenvalue problem requires that $S$ be positive-definite --- in the extreme case when $\hat{S}$ is rank-deficient, the maximum value of \cref{eq:geneig} is not defined and we get the matrix equivalent of a division by zero.

The most popular class of stabilized covariance matrix estimators are called shrinkage estimators, and take the form
\begin{align}
\hat{\mathbf{\Sigma}} = (1-\gamma)\mathbf{S} + \gamma  (\mathrm{Tr}(\mathbf{S}) / m) \mathbf{I},
\end{align}
for some positive constant $\gamma$. The interpretation of this expression is that the shrunk covariance matrix is a convex combination of two estimators, the (low bias, but high variance) sample covariance matrix, and the (high bias, but low variance) estimator that assumes all basis functions have identical variances and zero covariance. An estimator of this form was first popularized by Ledoit and Wolf in the context of Markowitz portfolio selection.\cite{ledoit2003improved, ledoit2004well, markowitz1952portfolio} Other shrinkage targets are possible beyond the scaled identity; we refer the reader to the excellent review by Sch\"{a}fer and Strimmer.\cite{schafer2005shrinkage}

The key insight of Ledoit and Wolf is that, under a Frobenius norm objective on the difference between the shrunk covariance matrix and the true covariance matrix, the asymptotically optimal value of the shrinkage constant, $\gamma$, can be estimated directly from $\mathbf{S}$, without knowing the true covariance matrix. Thus, no extra tunable parameters need to be added to the algorithm, which is important for usability.

Further improvements to the Ledoit-Wolf (LW) estimator  were made by Chen, Wiesel, and Hero III.\cite{chen2009shrinkage} First, using the Rao-Blackwell theorem,\cite{casella1996rao} they produced a more accurate Rao-Blackwellized Ledoit-Wolf (RBLW) estimator for the optimal shrinkage constant that dominates the LW estimator. In addition, unlike the LW estimator, the RBLW estimator can be computed even more efficiently and essentially requires no significant computational work beyond the calculation of the sample covariance matrix, $\mathbf{S}$. The expression for the RBLW-optimal shrinkage constant, $\gamma$, is
\begin{align}
    \gamma = \min(\alpha, \beta/U),
\end{align}
where $\alpha$, $\beta$, and $U$ are given by
\begin{align}
    \alpha &= \frac{N-2}{N(N+2)} \\
    \beta &= \frac{(m+1)N - 2}{N(N+2)} \\
    U &= \frac{m\Tr(\mathbf{S}^2)}{\Tr^2(\mathbf{S})} - 1.
\end{align}

We recommend this RBLW estimator for $\mathbf{\Sigma}$ for use with both tICA and sparse tICA.

\section{Projection of point onto an ellipsoid}
\label{appendix:projection}

Here we discuss our method for projecting a point in $\mathbb{R}^N$ onto an ellipsoid, following Kiseliov.\cite{kiseliov1994algorithms} Given a point $\mathbf{a}$ outside the ellipsoid and a positive definite matrix $\mathbf{\Sigma}$, the problem can be written as:
\begin{align}
\label{appendix:eq:ellipsoid}
\mathbf{z}^* = \begin{aligned}
&\argmin_\mathbf{z}\hspace{2em} ||\mathbf{z}-\mathbf{a}||^2 \\
&\text{subject to }\hspace{2em} \mathbf{z}^T \mathbf{\Sigma} \mathbf{z} \leq 1.\end{aligned}
\end{align}
Because, for our purposes, it will be necessary to solve the problem many times for different values of $\mathbf{a}$ with the same value of $\mathbf{\Sigma}$, it will be advantageous to consider any possible pre-processing of $\mathbf{\Sigma}$ that will speed up the calculation for each $\mathbf{a}$.

For the nontrivial case in which the point $\mathbf{a}$ lies outside the ellipsoid, the solution is on the border of the ellipsoid, ${\mathbf{z}^*}^T \mathbf{\Sigma} \mathbf{z}^* = 1$, so we address only the equality. First, consider the Lagrangian, $L$,
\begin{align}
L = ||\mathbf{z}-\mathbf{a}||^2 + \mu(\mathbf{z}^T\mathbf{\Sigma}\mathbf{z}-1).
\end{align}

The solution to \cref{appendix:eq:ellipsoid} satisfies the condition $\nabla L = 0$, yielding
\begin{align}
\mathbf{z}^* = (\mathbf{I}_n + \mu^* \mathbf{\Sigma})^{-1} \mathbf{a}.
\end{align}

The value of the Lagrange multiplier at the solution, $\mu^*$, must be determined to ensure that the constraint is satisfied. This requires solving the scalar equation $G(\mu)=0$, where $G(\mu)$ is defined as
\begin{align}
G(\mu) &= \mathbf{z}^*(\mu)^T \mathbf{\Sigma} \mathbf{z}^*(\mu) - 1, \\
\mathbf{z}^*(\mu) &= (\mathbf{I}_n + \mu \mathbf{\Sigma})^{-1} \mathbf{a}. \label{eq:xstarmu}
\end{align}

We solve for the root of $G$ using Newton's method, which requires computing $G$ and $G' = dG/d\mu$. Assuming that the eigendecomposition of $\mathbf{\Sigma}$ has been precomputed, $\mathbf{\Sigma}=\mathbf{V}\mathbf{D}(\mathbf{w})\mathbf{V}^T$, applying the Woodbury matrix identity shows that $G$ and $G'$ can be computed in linear time, without explicitly inverting any matrices or solving any linear systems, as \cref{eq:xstarmu} suggests might be necessary,
\begin{align}
\mathbf{z}^*(\mu) &= (\mathbf{I}_n + \mu \mathbf{\Sigma})^{-1} a \\
&= (\mathbf{I}_n + \mu \mathbf{V}\mathbf{D}(\mathbf{w})\mathbf{V}^T)^{-1} \mathbf{a} \\
&= (\mathbf{V} (\mathbf{I}_n + \mu \mathbf{D}(\mathbf{w}))\mathbf{V}^T)^{-1} \mathbf{a} \\
&= \mathbf{V}\mathbf{D}(\mathbf{e})\mathbf{V}^T a,
\end{align}
where $e_i = (\mu w_i +1)^{-1}$. Then, expanding $G(\mu)$, we have
\begin{align}
G(\mu) &= \mathbf{z}^*(\mu)^T \mathbf{\Sigma} \mathbf{z}^*(\mu) - 1 \\
&= (\mathbf{V}\mathbf{D}(\mathbf{e})\mathbf{V}^T \mathbf{a})^T \mathbf{V}\mathbf{D}(\mathbf{w})\mathbf{V}^T \mathbf{V}\mathbf{D}(\mathbf{e})\mathbf{V}^T \mathbf{a} - 1 \\
&= \mathbf{a}^T\mathbf{V} \mathbf{D}(\mathbf{f}) \mathbf{V}^T\mathbf{a} - 1,
\end{align}
where $f_i$ = $w_i e_i^2 = w_i/(\mu w_i + 1)^2$. The derivative required for Newton's method, $dG/d\mu$, is then very simple to calculate.

This algorithm is summarized in \cref{alg:2}. The quadratic convergence of Newton's method and low per-step work makes this preferable to alternatives such as the Lin-Han method.\cite{dai2006fast}

\begin{algorithm}[H]
\begin{algorithmic}
\Require $ \mathbf{a} \in \mathbb{R}^n, \mathbf{\Sigma} \in \mathbb{S}_{++}^n$
\State $\mathbf{w}, \mathbf{V} \gets \mathrm{eigs}(\mathbf{\Sigma})$ \Comment{Compute eigenvalues and eigenvectors}
\State $\mathbf{c} \gets \mathbf{V}^T\mathbf{a}$
\If{$\mathbf{c}^T\mathbf{D}(\mathbf{w})\mathbf{c} \leq 1$}
\State \Return $\mathbf{a}$ \Comment{Trivial if $\mathbf{a}$ is inside the set}
\Else
    \State $\mu^{(0)} \gets 1$
    \While{not converged} \Comment{Newton's method}
    \State $G^{(k)} \gets -1 + \sum_{i=1}^n c_i^2 w_i / (\mu^{(k)}w_i + 1)^2$
    \State $G'^{(k)} \gets -2\sum_{i=1}^n c_i^2 w_i^2 / (\mu^{(k)}w_i + 1)^3$
    \State $\mu^{(k+1)} \gets \mu^{(k)} - G^{(k)}/G'^{(k)}$
    \EndWhile
    \State $e_i \gets (\mu^{(k)}w_i + 1)^{-1}$
    \State \Return $\mathbf{V}\mathbf{D}(\mathbf{e})\mathbf{c}$
\EndIf
\caption{\label{alg:2} Projection of a point onto an ellipsoid}
\end{algorithmic}
\end{algorithm}

\section{Runtime performance}
\begin{figure}
    \includegraphics[width=0.5\textwidth]{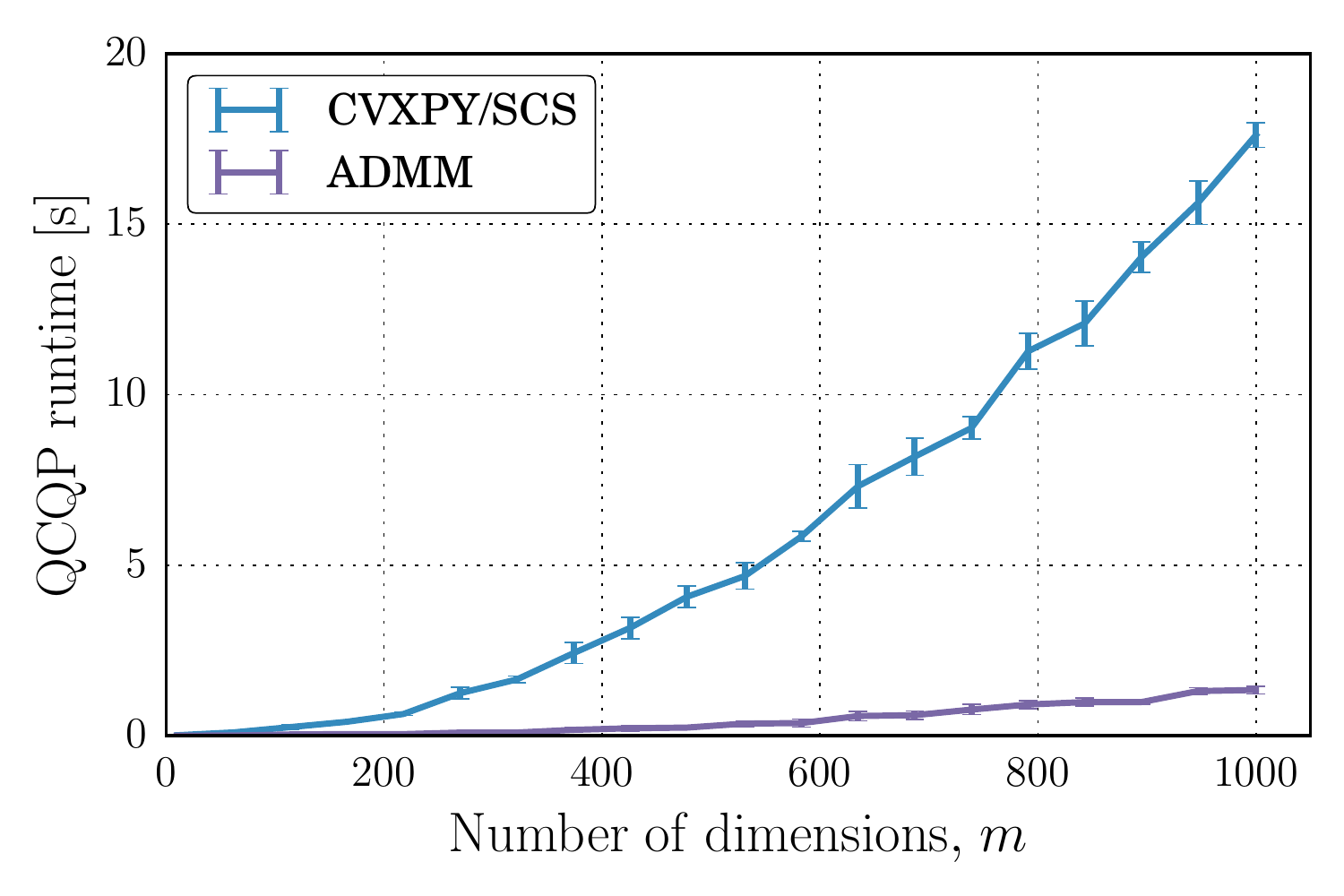}
    \caption{\label{fig:timing} Comparison of the runtime of our specialized QCQP solver and a generic solver using CVXPY and SCS.\cite{cvxpy, o2013operator} We observe a speedup of approximately one order of magnitude. Efficient warm-starting of the QCQP in \cref{alg:1} yields further improvements in runtime. Error bars indicate standard deviations over 5 replicates.}
\end{figure}

In addition to our ADMM-based solver, we implemented the sparse tICA algorithm using CVXPY and the off-the-shelf SCS solver to solve the QCQP.\cite{cvxpy, o2013operator} In \cref{fig:timing}, we compare the runtime of these two approaches. For this comparison, we randomly generated the matrix $\mathbf{\Sigma}$ from a Wishart distribution with $m$  degrees of freedom and an identity scale matrix, and initialized the ADMM solver from a vector, $\mathbf{x}$, with elements drawn from the standard normal distribution. The error bars indicate standard deviations over 5 replicates. The timings were performed on a Mid 2014 Apple Macbook Pro laptop.

We see generally that our solver is roughly an order of magnitude faster on the QCQP than CVXPY with SCS. Our sparse tICA implementation, however, is also able to efficiently warm-start, because the vectors $\mathbf{w}$ and $\mathbf{b}$ also converge during the outer iteration of \cref{alg:1}. Because of this, we find that when we substitute in the off-the-shelf solver to \cref{alg:1}, the speedup achieved by our ADMM approach is even more substantial. For example, while converging the first sparse tICA solution with $m=500$ using our ADMM implementation takes on the order of 0.1 seconds, the same optimization takes approximately 7 minutes using the off-the-shelf solver.

\section*{References}
\bibliography{references}

\end{document}